\documentclass[num-refs]{wiley-article}

\usepackage{siunitx}
\usepackage{caption}
\usepackage{longtable} 
\usepackage{pdflscape} 
\usepackage{multirow}
\usepackage{xcolor}
\usepackage[hidelinks]{hyperref}

\papertype{Original Article}
\title{Impact of predictor measurement heterogeneity across settings on performance of prediction models: a measurement error perspective}

\author[1]{K. Luijken}
\author[1,2]{R.H.H. Groenwold}
\author[2,3]{B. van Calster}
\author[2,4]{E.W. Steyerberg}
\author[1]{M. van Smeden}
\affil[1]{Department of Clinical Epidemiology, LUMC, Leiden, the Netherlands}
\affil[2]{Department of Biomedical Data Sciences, LUMC, Leiden, the Netherlands}
\affil[3]{Department of Development and Regeneration, University of Leuven, Leuven, Belgium}
\affil[4]{Department of Public Health, Erasmus MC, Rotterdam, the Netherlands}
\corraddress{Kim Luijken, Department of Clinical Epidemiology, LUMC, Leiden, the Netherlands}
\corremail{K.Luijken@lumc.nl}
\fundinginfo{RG was supported by the Netherlands Organisation for Scientific Research (ZonMW, project 917.16.430). ES was supported through a Patient-Centered Outcomes Research Institute (PCORI) Award (ME-1606-35555).}

\begin{document}
\maketitle

\begin{abstract}
\small{It is widely acknowledged that the predictive performance of clinical prediction models should be studied in patients that were not part of the data in which the model was derived. Out-of-sample performance can be hampered when predictors are measured differently at derivation and external validation. This may occur, for instance, when predictors are measured using different measurement protocols or when tests are produced by different manufacturers. Although such heterogeneity in predictor measurement between deriviation and validation data is common, the impact on the out-of-sample performance is not well studied. Using analytical and simulation approaches, we examined out-of-sample performance of prediction models under various scenarios of heterogeneous predictor measurement. These scenarios were defined and clarified using an established taxonomy of measurement error models. The results of our simulations indicate that predictor measurement heterogeneity can induce miscalibration of prediction and affects discrimination and overall predictive accuracy, to extents that the prediction model may no longer be considered clinically useful. The measurement error taxonomy was found to be helpful in identifying and predicting effects of heterogeneous predictor measurements between settings of prediction model derivation and validation. Our work indicates that homogeneity of measurement strategies across settings is of paramount importance in prediction research.
\keywords{Prediction model, measurement error, measurement heterogeneity, external validation, calibration, discrimination, Brier score}}
\end{abstract}

\newpage
\section{Introduction}
Prediction models have an important role in contemporary medicine by providing probabilistic predictions of diagnosis or prognosis \cite{stey2008}. Prediction models need to provide accurate and reliable predictions for patients that were not part of the dataset in which the model was derived (i.e., derivation set) \cite{altman2000we}. The ability of a prediction model to predict in future patients (i.e., out-of-sample) can be evaluated in an external validation study. While out-of-sample predictive performance is in general expected to be lower than performance estimated at derivation \cite{stey2008}, large discrepancies are often contributed to suboptimal modeling stategies in the derivation of the model \cite{collins2016sample,steyerberg2004validation,steyerberg2017poor} and differences between patient characteristics in derivation and validation samples \cite{vergouwe2010external, debray2013framework}. \par
Another potential source of limited out-of-sample performance is when predictors are measured differently at derivation than at (external) validation. This may occur, for instance, when predictors are categorized using different cut-off values or when predictors are based on diagnostic tests that were produced by different manufacturers (see Table \ref{Examples} for examples). Although some studies have mentioned that such heterogeneity in predictor measurements might hamper out-of-sample model performance (e.g.,\cite{collins2015transparent,wynants2013screening}), effects of measurement heterogeneity in prediction studies have received little attention. Particularly, its impact on predictive performance has not been formally quantified.\par 
In this study, we investigate the out-of-sample performance of a clinical prediction model in situations where predictor measurement strategies at the model derivation stage differed from measurement strategies at the model validation stage. The different scenarios of heterogeneous predictor measurement were defined using a well-known taxonomy of measurement error models, described by e.g. Keogh et al. \cite{keogh2014toolkit}. We varied the degree of measurement error in the derivation data and validation data to recreate qualitative differences in the predictor measurement structures across settings. Hence, the measurement error perspective serves as a framework to define predictor measurement heterogeneity. We focus on logistic regression, since this model is widely applied in clinical prediction research \cite{stey2010assessing}.\par
This paper is structured as follows. In Section 2, we define the measurement error models used to describe scenarios of measurement heterogeneity. In Section 3, we derive analytical expressions to identify and predict effects of measurement error on in-sample predictive performance. In Section 4, we illustrate the effects of measurement heterogeneity across settings  on predictive performance in large sample simulations and contrast these to the impact of measurement error within the derivation setting. In Section 5, we present an extensive set of Monte Carlo simulations in finite samples to examine the impact of measurement heterogeneity on out-of-sample predictive performance. We end with discussing the implications of our findings in Section 6. 

\section{Expressing measurement heterogeneity in terms of measurement error models} 
Consider a random sample of $N$ independent individuals $i = 1,\ldots,N$. Let $Y$ be a binary response variable with values $y_i \in \{0,1\}$. We define a logistic regression model for estimating the probability that $Y = 1$ given values of a set of $P$ continuous predictor variables, $\bm{X} = \{X_1,\ldots, X_P$\}. The probability of observing an event ($Y = 1$) given the predictors, $\pi_{i} = P(Y_i = 1 \vert \bm{X}_{i})$, is defined as
\begin{equation*}
\label{logistic prob}
\pi_{i} = \frac{1}{1+\mbox{exp}(-(\alpha+\bm{\beta}^T \bm{x}_{i}))},
\end{equation*}
where $\alpha$ is an intercept (scalar),  $\bm{\beta}$ is a $P$-dimensional vector of regression coefficients. \par
For simplicity of presentation, we consider a single vector $X 	\subset \bm{X}$. To distinguish different measurements of the same predictor, we denote an exact measurement of the predictor (e.g. bodyweight measured on a scale) by $X$ and a pragmatic measurement (e.g. self-reported weight) by $W$. In most measurement error literature, $X$ denotes an error-free true value and $W$ denotes an observed error-prone version of $X$ \cite{carroll2006measurement}. However, for prediction purposes, it is hardly ever feasible (or even undesirable) to obtain error-free measurements in clinical practice, and hence we use the terms exact measurement for $X$ and pragmatic measurement for $W$. The connection between $X$ and $W$ can be formally defined using measurement error models. We define a general model of measurement heterogeneity for continuous predictors in line with existing measurement error literature \cite{keogh2014toolkit,carroll2006measurement}. Assuming that the relation between $X$ and $W$ is linear and additive, the association between $W$ and $X$ can be described as
\begin{align}
\label{General MEmodel}
\mathbb{E}(W \vert Y = y) &= \psi_{Y=y} + \theta_{Y=y}\mathbb{E}(X)  + \epsilon_{Y=y}, \\
\text{Var}(W \vert Y = y) &= \theta_{Y=y}^2\sigma_X^2  + \sigma_{\epsilon_{Y=y}}^2,\nonumber 
\end{align}
where $\epsilon_{Y = y} \sim \mathcal{N}(0,\sigma_{\epsilon_Y=y}^2)$ and all parameters may depend on the value of $Y$, indicating that measurements can differ between individuals in which the outcome is observed (cases) and individuals in which the outcome is not observed (non-cases). The parameter $\psi$ reflects the mean difference between $X$ and $W \vert Y = y$, $\theta$ indicates the linear association between measurement $W \vert Y = y$ and $X$, and $\sigma_{\epsilon}^2$ reflects variance introduced by random deviations in the measurement process, where a larger $\sigma_{\epsilon}^2$ indicates that the measurement $W$ is less precise. The term \textit{measurement error} applies to situations where both an exact measurement and a pragmatic measurement of a predictor are available within a setting (e.g., the derivation set), and thus where the parameters $\psi$, $\theta$ and $\sigma_{\epsilon}^2$ define the degree of measurement error in $W$ with respect to $X$. The term \textit{measurement heterogeneity} refers to situations where the same predictor is measured heterogeneously across settings of derivation and validation. The most precise measurement (whether available at derivation or validation) corresponds to $X$ and the parameters $\psi$, $\theta$ and $\sigma_{\epsilon}^2$ define the degree of heterogeneity between $X$ and $W$. We now consider three types of measurement error models that are particular forms of Equation (\ref{General MEmodel}), based on which we specify both within-sample measurement error and measurement heterogeneity across settings. 
\paragraph{Random measurement error model} 
Under $\psi=0$ and $\theta=1$, Equation (\ref{General MEmodel}) reduces to the following model:
\begin{equation}
\label{Random ME}
\mathbb{E}(W) = \mathbb{E}(X) + \epsilon,
\end{equation}
where $\epsilon \sim \mathcal{N}(0,\sigma_{\epsilon}^2)$ is independent of $X$ and $Y$. This is referred to as the random or classical measurement error model \cite{keogh2014toolkit,carroll2006measurement}. $W$ is a mean-unbiased measurement of $X$, since $\mathbb{E}(W \vert Y) = \nolinebreak \mathbb{E}(W) = \nolinebreak \mathbb{E}(X)$. An example of a predictor measurement corresponding to the random measurement error model is reading body weight from the same scale. Each reading, the value may deviate slightly upwards or downwards, resulting in random deviations. Variation in the size of these deviations across settings due to precision of the available scales is an example of random measurement heterogeneity.
\paragraph{Systematic measurement error model}
When $\psi\neq0$ and/or $\theta\neq1$, yet when $\psi$ and $\theta$ have the same values for cases and non-cases, predictor measurements correspond to a systematic measurement error model \cite{keogh2014toolkit}. The systematic measurement error model is defined as
\begin{equation}
\label{Systematic ME}
\mathbb{E}(W) = \psi+ \theta \mathbb{E}(X)+ \epsilon,
\end{equation}
where $\epsilon \sim \mathcal{N}(0,\sigma_{\epsilon}^2)$ is independent of $X$ and $Y$. It follows that $W$ is no longer a mean-unbiased measurement of X ($\mathbb{E}(W) \neq \mathbb{E}(X)$). Systematic measurement heterogeneity may occur, for example, when a blood glucose monitor is replaced by a monitor from a different manufacturer that is calibrated differently. The switch in measurement instrument may introduce a shift by a constant in the measured predictor values, i.e. a change in $\psi$ (additive systematic measurement error). Furthermore, observed values may depend on the actual value of a predictor, where $\theta$ represents linear dependencies between $X$ and $W$. For instance, values of self-reported weight may be underreported, especially by individuals with a higher actual weight, i.e. $\theta < 1$ (multiplicative systematic measurement error). The size of $\psi$ and $\theta$ can differ across settings, for example when weight is measured using a scale in one setting (e.g. $\theta$ might be close to $1$) and as a self-reported value in another setting (e.g. $\theta$ might deviate from $1$), which would result in systematic measurement heterogeneity.
\paragraph{Differential measurement error model}
In case measurement procedures differ between cases and non-cases, i.e. when $\psi_1 \neq \psi_0 \text{ and/or } \theta_1 \neq \theta_0 $ \text{and/or} $\sigma_{\epsilon1}^2 \neq \sigma_{\epsilon0}^2$, the measurements can be described by Equation (\ref{General MEmodel}) above, also referred to as differential measurement error \cite{keogh2014toolkit}. Differential measurement of predictors is conceivable in settings where assessment of predictors are done in an unblinded fashion, such as case-control studies \cite{white2011measurement}. For example, when patient history is collected after observing the outcome event, cases may be more likely to recall health information prior to the outcome event than non-cases, also known as recall bias \cite{sackett1979bias}. This may for example lead to over-reporting in cases, i.e. $\psi_1 > \psi_0$, a stronger association between reported and actual predictor values, i.e. $\theta_1 > \theta_0$, or more precise predictor measurements, i.e. $\sigma_{\epsilon_1}^2 < \sigma_{\epsilon_0}^2$, in cases than in non-cases. Prospective differential measurement error may occur when a prediction model influences the way that predictors are measured in clinical practice. After clinical uptake of a prediction model, physicians may measure predictors differently in patients in whom they suspect the outcome of interest (potential future cases), guided by the knowledge that these particular predictors are of importance. For example, in these patients, body weight may be measured using a scale, whereas the prediction model may have been derived from self-reported measurements of body weight, introducing a difference between measurement procedures of (potential) cases and non-cases (i.e., differential measurement error), as well as a difference in measurement strategy between derivation and application setting (i.e., differential measurement heterogeneity).

\section{Predictive performance under within-sample measurement error}
In this section, we define analytical expressions that indicate how substituting an exact predictor measurement, $X$, with a pragmatic predictor measurement, $W$, affects apparent predictive performance in the situation where both measurements $X$ and $W$ are available in the derivation sample of a prediction model. For brevity, we will evaluate a single-predictor model. Expressions of in-sample predictive performance under random measurement error were previously derived by Khudyakov and colleagues for a probit prediction model \cite{khudyakov2015impact}. The current paper extends these expressions to a logistic regression model. We measure predictive performance by the concordance-statistic (c-statistic) and Brier score, measuring discrimination and overall accuracy, respectively. Effects on calibration will be evaluated in the next sections. We will discuss expressions in terms of sample realizations, that is, realizations $y_i$, $x_i$ and $w_i$. In the following, let $\bar{x} = \frac{1}{n} \sum_{i=1}^{n} (x_i \vert y_i)$ and $s_x^2$ denote the sample mean and variance of $x$, let $\bar{w} = \frac{1}{n} \sum_{i=1}^{n} (w_i \vert y_i)$ and $s_w^2$ denote the sample mean and variance of $w$, and let $n_1$ and $n_0$ denote the number of cases and non-cases in the sample, respectively. 

\subsection{C-statistic}
To examine the discriminatory performance, we make use of the c-statistic, a rank-order statistic that typically ranges from 0.5 (no discrimination) to 1 (perfect discrimination) and is equal to the area under the receiver operating characteristic (ROC) curve for a binary outcome \cite{stey2010assessing}. Consider a data generating model relating response variable $Y$ to $X$ by a logit link function, where $X \vert Y \sim \mathcal{N}(\mu_Y,\sigma_Y^2)$ (binormality). Let $\bar{x_1}= \frac{1}{n_1} \sum_{i=1}^{n_1} (x_i \vert y_i=1)$ denote the sample mean of $x$ for cases, let $\bar{x_0} = \frac{1}{n_0} \sum_{i=1}^{n_0} (x_i \vert y_i=0)$ denote the sample mean of $x$ for non-cases, and let $s_{x1}^{2}+s_{x0}^{2}$ denote the total variance of $x$. Let $\Phi$ denote the cumulative distribution function of the standard normal distribution. Following Austin and Steyerberg\cite{austin2012interpreting}, the c-statistic is approximated by 
\begin{equation*}
\label{c-statistic}
AUC_x = \Phi\left(\frac{\bar{x_1}-\bar{x_0}}{\sqrt{s_{x1}^{2}+s_{x0}^{2}}} \right).\\
\end{equation*}
\newline
Alternatively, for $w$, let $\bar{w_1} = \frac{1}{n_1} \sum_{i=1}^{n_1} (w_i \vert y_i=1)$ and $\bar{w_0} = \frac{1}{n_0} \sum_{i=1}^{n_0} (w_i \vert y_i=0)$ denote the sample means of $w$ for cases and non-cases, respectively, and let $s_{w1}^{2}+s_{w0}^{2}$ denote the total variance of $w$. The c-statistic of a binary logistic regression model of the predictor $w$ is then given by:
\begin{equation}AUC_w = \Phi \left(\frac{\bar{w_1}-\bar{w_0}}{\sqrt{s_{w1}^{2}+s_{w0}^{2}}} \right).\\
\label{cstatME}
\end{equation}
Under the general measurement error model (Equation \ref{General MEmodel}),
\begin{align*}
\bar{w_0} &= \psi_0 + \theta_0\bar{x_0},\\
\bar{w_1} &= \psi_1 + \theta_1\bar{x_1},\\
s_{w0}^{2} &= s_{x0}^2\theta_0^2 + s_{\epsilon_0}^2,\\
s_{w1}^{2} &= s_{x1}^2\theta_1^2 + s_{\epsilon_1}^2.
\end{align*}
The impact of measurement error on the c-statistic can now be expressed as
\begin{align}
\Delta AUC &= AUC_w - AUC_x \nonumber \\
&= \Phi \left(\frac{(\psi_1 + \theta_1\bar{x_1})-(\psi_0 + \theta_0\bar{x_0})}{\sqrt{s_{x1}^2\theta_1^2 + s_{\epsilon_1}^2+s_{x0}^2\theta_0^2 + s_{\epsilon_0}^2}} \right) - \Phi\left(\frac{\bar{x_1}-\bar{x_0}}{\sqrt{s_{x1}^{2}+s_{x0}^{2}}} \right),\label{cstatMEdif}
\end{align}
where a $\Delta AUC < 0$ indicates that the model has less discriminatory power when $w$ is used instead of $x$. Equations (\ref{cstatME}) and (\ref{cstatMEdif}) indicate that the expected impact of substituting $x$ by $w$ in prediction model development has the following consequences. In case of random measurement error in $w$, it can be expected that the model fitted on $w$ has a lower c-statistic and $\Delta AUC < 0$. In case of systematic measurement error in $w$, the c-statistic is not affected beyond random measurement error. Differential measurement error can affect model discrimination in both directions. For example, when observed measurements $w$ are systematically shifted further from $x$ in cases, i.e. when $\psi_1 > \psi_0$ and $\theta_1 = \theta_0 = 1$, and when the difference in mean predictor values between cases and non-cases in $x$ is positive, i.e. $\bar{x_1} > \bar{x_0}$ and $AUC_x > 0.5$, the mean difference in predictor values between cases and non-cases, $\bar{w_1}-\bar{w_0}$, increases, enlarging the discriminatory power of the model, i.e. $\Delta AUC > 0$. Additional random measurement error affects the c-statistic irrespective of whether the error is differential or not.

\subsection{Brier score}
As a measure of overall predictive accuracy we evaluate the Brier score, which is a proper scoring rule that indicates the distance between predicted and observed outcomes. The Brier score is calculated by \cite{brier1950verification}
\begin{equation}
\label{Empirical Brier}
BS(x) = \frac{1}{n}\sum_{i=1}^n(y_i - \hat{\pi}(x_i))^2,
\end{equation}
where $\hat{\pi}(x_i) = (1+\mbox{exp}(-(\hat{\alpha}_{x}+ \hat{\beta}_{x}x_i)))^{-1}$ and a lower Brier score indicates higher accuracy of predictions. Following \cite{blattenberger1985separating} and \cite{spiegelhalter1986probabilistic}, the Brier score can be decomposed into  
\begin{equation}
\label{Decomposed Brier}
BS(x) = \frac{1}{n}\sum_{i=1}^n(y_i - \hat{\pi}(x_i))(1-2\hat{\pi}(x_i)) + \frac{1}{n}\sum_{i=1}^n\hat{\pi}(x_i)(1-\hat{\pi}(x_i)),
\end{equation}
resulting in a calibration component, $(y_i - \hat{\pi}(x_i))(1-2\hat{\pi}(x_i))$, and a refinement component, $\hat{\pi}(x_i)(1-\hat{\pi}(x_i))$. As Spiegelhalter already noted \cite{spiegelhalter1986probabilistic}, the calibration component has an expectation of $0$ under the null hypothesis of perfect calibration, that is $\mathbb{E}_0(Y_i) = \hat{\pi}(x_i)$, and the expected Brier score can be expressed by the refinement term in Equation (\ref{Decomposed Brier}), that is $\mathbb{E}_0(BS(x)) = \frac{1}{n}\sum_{i=1}^n\hat{\pi}(x_i)(1-\hat{\pi}(x_i))$. Consequently, the impact of within-sample measurement error on the Brier score of a maximum likelihood model in the derivation set can be expressed as 
\begin{equation}
\label{Difference Brier}
\mathbb{E}_0(\Delta BS) = \frac{1}{n}\sum_{i=1}^n\hat{\pi}(w_i)(1-\hat{\pi}(w_i)) - \frac{1}{n}\sum_{i=1}^n\hat{\pi}(x_i)(1-\hat{\pi}(x_i)),
\end{equation}
where
\begin{equation*}
\hat{\pi}(w_i) = \frac{1}{1+\text{exp}(-(\hat{\alpha}_{w}+ \hat{\beta}_{w}(\psi_{Y=y} + x_i\theta_{Y=y} + \epsilon_{Y=y}))},
\end{equation*}
and where a $\mathbb{E}_0(\Delta BS) > 0$ indicates that substituting $x$ with $w$ yields less accurate predictions. Realistically, however, a model is hardly ever perfectly calibrated (see \cite{van2016calibration} for an in-depth discussion of levels of calibration of prediction models). A maximum likelihood estimate of a logistic regression model attains 'weak calibration' in its derivation sample by definition, meaning that no systematic overfitting or underfitting and/or overestimation or underestimation of risks occurs. In the remaining of this paper we use the term 'calibration' instead of 'weak calibration' and use the term 'Brier score' to refer to the decomposed empirical Brier score in Equation (\ref{Decomposed Brier}).\par
Expression (\ref{Difference Brier}) indicates that substituting $x$ with $w$ in a perfectly specified model has the following consequences. When the association between $w$ and outcome $y$ is weaker than the association between $x$ and $y$, a prediction model based on $w$ provides less extreme predicted probabilities. This results in a larger refinement term for $w$, i.e. $\frac{1}{n}\sum_{i=1}^n\hat{\pi}(w_i)(1-\hat{\pi}(w_i))$ is larger, and in a positive $\mathbb{E}_0(\Delta BS)$ and hence lower accuracy.

\section{Measurement error versus measurement heterogeneity}
The expressions of predictive performance under measurement error indicate that more erroneous predictor measurements lead to less apparent discriminatory power and accuracy. However, these results cannot be generalized directly to effects of measurement error on out-of-sample performance of prediction models. We use the measurement error model taxonomy to explore how heterogeneity in measurement structures affects out-of-sample performance. Rather than distinguishing error-free and error-prone predictor measurements, the measurement error models now express deviations from homogeneity of measurements across settings.\par
A direct comparison of effects of measurement error and effects of measurement heterogeneity on predictive performance can be found in Figure 1 and 2, which illustrate large-sample (N $= 1,000,000$) properties of predictive performance measures. Effects of measurement error are illustrated by comparing in-sample predictive performance measures of a prediction model that is first estimated based on $x$ and subsequently estimated based on $w$, where the latter contains increasing measurement error. Effects of measurement heterogeneity are illustrated by comparing out-of-sample predictive performance measures of a prediction model that is transported across settings with different predictor measurement structures. We explored three settings: (i) $x$ is available at derivation and $w$ is available at validation, (ii) $w$ is available at both derivation and validation, and (iii) $w$ is available at derivation and $x$ is available at validation. In other words, this section illustrates the impact of measurement error and measurement heterogeneity as an isolated factor by evaluating the same population at both derivation and validation, and only varying the predictor measurement structures. For the purpose of demonstration, we focus on random measurement error and -heterogeneity and provide further analyses in the next section.\par
Additional to the c-statistic and Brier score, we evaluate calibration as a measure of predictive performance. In logistic regression, calibration can be determined using a re-calibration model, where the observed outcomes in validation data, $y_V$, are regressed on a linear predictor (lp) \cite{cox1958two}. This linear predictor is obtained by combining the regression coefficients estimated from the derivation data, $\hat{\alpha}_D$ and $\hat{\beta}_D$, with the predictor values in the validation data, $x_{iV}$. The recalibration model is defined as\cite{stey2008} \hspace{-0.6mm}:
\begin{equation*}
\label{recalibration}
\text{logit}(y_V) = a + b\times\text{lp},
\end{equation*}
where $\text{lp} = \hat{\alpha}_D + \hat{\beta}_Dx_{iV}$ and $b$ represents the calibration slope. A calibration slope $b = 1$ indicates perfect calibration. A calibration slope $b < 1$ indicates that predicted probabilities are too extreme compared to observed probabilities, which is often found in situations of 'statistical overfitting' \cite{stey2008,van2016calibration}. A calibration slope $b > 1$ indicates that the provided predicted probabilities are too close to the outcome incidence, also referred to as 'statistical underfitting'. Additional to the calibration slope, we evaluated the difference between the average observed event rate and the mean predicted event rate (i.e. calibration-in-the-large, which can be computed as the intercept of the recalibration model while using an offset for the linear predictor, i.e., $a \vert b = 1$ ).\cite{stey2008}\par
In situations of within-sample measurement error, i.e. in the re-estimated model, all calibration plots showed a calibration slope equal to $b=1$, indicating perfect apparent calibration (Figure 1A-C). The apparent c-statistic and Brier score improved with decreasing random measurement error. In case of measurement heterogeneity across samples, i.e. in the transported model, similar changes in the c-statistic and Brier score were found. However, heterogeneous measurements led to a calibration slope $b \neq 1$, indicating that predictions were no longer valid (Figure 1D and 1F). When measurements at validation were less precise than at derivation, the calibration slope was $b < 1$, similar to statistical overfitting. When measurements at validation were more precise than at derivation, the calibration slope was $b > 1$, similar to statistical underfitting. More elaborate illustrations of the impact of measurement heterogeneity in large sample simulations, including effects of systematic and differential measurement heterogeneity, can be found in Appendix 1.\par
Although the total Brier score did not differ substantially between the re-estimated and transported model, examination of the large sample properties of the decomposed Brier score (Equation \ref{Decomposed Brier}) indicated differences in the components between the procedures (Figure 2). In the re-estimated model, the calibration term equaled zero, and the total Brier score equaled the refinement term (Figure 2A). The Brier score increased with increasing random measurement error, indicating that accuracy decreased. In the transported model, changes in the refinement term were counterbalanced by changes in the calibration term. For example, when measurements at validation were less precise than at derivation, the spread in predicted probabilities increased (refinement term in Figure 2b decreased). A decrease in the refinement term under perfect calibration would indicate that overall accuracy of the model is improving, as predicted probabilities are closer to $0$ or $1$. However, in the transported model this improvement was counterbalanced by a calibration term larger than zero, which indicates that predicted probabilities were too extreme compared to observed probabilities (Figure 2B). \par
Figure 1 and 2 illustrate that miscalibration is not introduced by measurement error per se, but rather by measurement heterogeneity across settings of derivation and validation. The discrepancy in calibration between model re-estimation and model transportation can be reduced to differences in the linear predictors of the recalibration models. In case of model re-estimation, the linear predictor is expressed by
\begin{equation}
\label{lpwd}
\text{lp}_{re-est} = \hat{\alpha}_{w(V)} + \hat{\beta}_{w(V)} w_{iV},
\end{equation} 
indicating that the parameters $\hat{\alpha}_{w(V)}$ and $\hat{\beta}_{w(V)}$ are estimated using the predictor values measured by strategy $w$ in the validation data. In the more realistic validation procedure in which the model is transported over different predictor measurement procedures, the linear predictor is expressed by
\begin{equation}
\label{lpwv}
\text{lp}_{transp} = \hat{\alpha}_{x(D)} + \hat{\beta}_{x(D)} w_{iV},
\end{equation} 
meaning that regression coefficients are estimated based on $x_{iD}$ and that the model is validated using $w_{iV}$. This distinction in recalibration models sheds a different light on previous research into effects of measurement error on predictive performance. Khudyakov and colleagues derived analytically that calibration in a derivation sample is not affected by measurement error \cite{khudyakov2015impact}. Since their findings are based on the assumption that the linear predictor is defined as in Equation (\ref{lpwd}), previous results on the impact of measurement error on predictive performance can be interpreted as effects on in-sample predictive performance \cite{khudyakov2015impact,rosella2012influence}.\par

\section{Predictive performance under measurement heterogeneity across settings}
General patterns of predictive performance under measurement heterogeneity were examined in a set of Monte Carlo simulations in finite samples to evaluate their behavior under sampling variability. Simulations were performed in R version 3.3.1. \cite{team2011r} and our code is accessible online (see \url{https://github.com/KLuijken/Prediction_Measurement_Heterogeneity_Predictor}). We studied the predictive performance of a single- and a two-predictor binary logistic regression model. For the latter, we evaluated situations in which both predictors were measured heterogeneously across settings as well as situations in which one of the predictors was measured similar over settings. The data for the single-predictor model were generated from 
\begin{align*}
\text{logit}(Y)  &= \text{log}(4) X,\\
\text{where } X &\sim \mathcal{N}(0,1). 
\end{align*}
\noindent The data for the two-predictor models were generated from 
\begin{align*}
\text{logit}(Y) &= \bm{\beta}^T \bm{X},\\
\text{where }\bm{X} &\sim \mathcal{N}\left(\begin{psmallmatrix}0\\0\end{psmallmatrix},\begin{psmallmatrix}1 & \rho_{X1X2}\\ \rho_{X1X2} & 1\end{psmallmatrix}\right). 
\end{align*}
The correlation between predictors, $\rho_{X1X2}$, varied with $0$, $0.5$ and $0.9$. Both the $\beta$-parameters in the two-predictor models have value $2.3$ in case $\rho_{X1X2} = 0$ or $\rho_{X1X2} = 0.5$, and have value $2.1$ in case $\rho_{X1X2} = 0.9$. We varied the values of the regression coefficients in order to keep the c-statistic of the data-generating models at an approximate value of $0.80$ and hence to compare predictive performance over models \cite{austin2012interpreting}. We recreated different measurement procedures of the predictors using different specifications of the general measurement error model (Equation \ref{General MEmodel}). In the derivation sample, measurements corresponded to the random measurement error model (Equation \ref{Random ME}), while in validation various measurement structures were recreated (see Table \ref{tab: factorial} for values of input parameters). All measurements contained at least some erroneous measurement variance to generate realistic scenarios. \par
In total, 432 scenarios were evaluated. For each scenario, a derivation sample ($n = 2,000$) and a validation sample ($n = 2,000$) were generated. We did not consider smaller sample sizes, since predictive performance measures are sensitive to statistical overfitting, which would complicate the interpretation of effects of measurement heterogeneity \cite{steyerberg2004validation,steyerberg2017poor}. The validation procedure was repeated 10,000 times for each simulation scenario. The number of events was around $1,000$ in each dataset, which exceeds the minimal requirement for validation studies \cite{vergouwe2005substantial,van2016calibration}. 

\paragraph{Simulation outcome measures}
The simulation outcome measures were the average c-statistic, calibration slope, calibration-in-the-large coefficient, and Brier score. The c-statistic was computed using the \texttt{somers2} function of the rms package \cite{harrell2013rms}. The calibration slope was computed by regressing the observed outcome in the validation dataset on the linear predictor, as defined in Equation (\ref{lpwd}). We evaluated calibration graphically by plotting loess calibriation curves and overlaying the plots of all 10,000 resamplings \cite{van2016calibration,austin2014graphical}. The calibration-in-the-large was computed as the intercept of the recalibration model, while using an offset for the linear predictor \cite{stey2008}. The empirical Brier score was computed using Equation (\ref{Empirical Brier}). Additionally, we evaluated in-sample predictive performance as a reference for effects on out-of-sample performance.

\subsection{Simulation results}
Identical measurement error structures at derivation and validation resulted in consistent predictive performance across settings. All out-of-sample measures of predictive performance were affected by measurement heterogeneity. Effects on predictive performance measures were largest in the single-predictor model (Table 3). The two-predictor model in which one of the predictors was measured consistently over settings (Figure 3) outperformed the model in which none of the predictors were measured consistently across settings (Figure 4). Inspection of calibration plots confirmed all patterns of miscalibration discussed below (Supplementary Figure 1). By and large, the impact of correlation between predictors on other parameters was minimal since the correlation structure was equal across compared settings, hence, we show combined results in the figures. 

\subsubsection{Random measurement heterogeneity}
When measurements were less precise at validation compared to derivation, i.e. when $\sigma_{\epsilon(D)}^2 < \sigma_{\epsilon(V)}^2$, the c-statistic decreased and Brier score increased at validation. In the single-predictor model, the c-statistic decreased from $0.75$ at derivation to $0.59 - 0.73$ at validation and the Brier score increased from $0.20$ at derivation to $0.23 - 0.28$ at validation (Table 3, bottom rows). Furthermore, the median calibration slope at validation was smaller than 1, ranging from $0.25 - 0.43$ in the single-predictor model. When measurements were more precise at validation compared to derivation, i.e. when $\sigma_{\epsilon(D)}^2 > \sigma_{\epsilon(V)}^2$, the c-statistic was increased, from $0.66$ to $0.68 - 0.78$ in the single-predictor model, and the Brier score was decreased, changing from $0.23$ to $0.20 - 0.24$ in the single-predictor model. However, the improved c-statistic and Brier score were accompanied by median calibration slopes greater than 1, ranging from $2.16 - 3.16$ in the single-predictor model (Table 3, top rows). Calibration-in-the-large was not affected by random measurement heterogeneity. Similar effects on predictive performance were observed for the two-predictor models, which are presented graphically in Figures 3 and 4.

\subsubsection{Systematic measurement heterogeneity}
When measurements at external validation changed by a constant compared to derivation, i.e. when $\psi_D = 0$ and $\psi_V = 0.25$, the risk on observing the outcome was systematically overestimated, which is reflected in the negative value for calibration-in-the-large coefficient (Table 3). Changes in $\psi$ had little effect on the calibration slope and Brier score, and no apparent effect on the c-statistic. Multiplicative systematic measurement heterogeneity, i.e. $\theta_D \neq \theta_V$, reinforced or counterbalanced effects of random measurement heterogeneity in the direction of the systematic measurement heterogeneity. When the association between $x$ and $w$ was relatively weak at validation, e.g. when $\theta_V = 0.5$, predictive performance deteriorated (black bars in Figures 3 and 4), whereas predictive performance improved when the association between $x$ and $w$ was relatively strong, e.g. when $\theta_V = 2.0$ (gray bars in Figures 3 and 4).

\subsubsection{Differential measurement heterogeneity}
We highlight four specific scenarios in which the single-predictor model was derived under differential random measurement error, i.e. $\sigma_{\epsilon1}^2 \neq \sigma_{\epsilon0}^2$, and validated using non-differential measurements, and vice versa (Table 4). Differential measurement led to miscalibration at external validation in all scenarios. The c-statistic and Brier score at validation slightly improved when cases were measured less precise at derivation or more precise at validation. For example, when cases were measured less precise at derivation, i.e. $\sigma_{\epsilon1(D)}^2 > \sigma_{\epsilon0(D)}^2$, the c-statistic increased from $0.66$ to $0.71$ at validation and the Brier score decreased from $0.23$ to $0.22$. However, the median calibration slope at validation was $1.86$.

\section{Discussion}
Heterogeneity of predictor measurements across settings can have a substantial impact on the out-of-sample performance of a prediction model. When predictor measurements are more precise at derivation compared to validation, model discrimination and accuracy at validation deteriorate, and the provided predicted probabilities are too extreme, similar to when a model is overfitted with respect to the derivation data. When predictor measurements are less precise at derivation compared to validation, discrimination and accuracy at validation tend to improve, but the provided predicted probabilities are too close to the outcome prevalence, similar to statistical underfitting. These key findings of our study are summarized in Table 5. The current study emphasizes that a prediction model not only concerns the algorithm relating predictors to the outcome, but also depends on the procedures by which model input is measured, i.e. qualitative differences in data collection.\par
Measurement error is commonly thought not to affect the validity of prediction models, based on the general idea that unbiased associations between predictor and outcome are no prerequisite in prediction studies \cite{carroll2006measurement}. By taking the measurement error perspective, our study revealed that prediction research requires consideration of variation in measurement procedures \textit{across} different settings of derivation and validation, rather than analyzing the amount of measurement error \textit{within} a study. A recent systematic review by Whittle and colleagues demonstrated that measurement error was not acknowledged in many prediction studies, and pointed out the need to investigate consequences of measurement error in prediction research \cite{whittle2018measurement}. An important starting point for this research following from our study is that the generalizability of prediction models depends on the transportability of measurement structures.\par
Specification of measurement heterogeneity can help to explain discrepancies in predictive performance between derivation and validation setting in a pragmatic way. The relatedness between derivation and validation samples is generally quantified in terms of similarity in person-characteristics (also referred to as "case-mix"), and regression coefficients \cite{stey2008}. Previously proposed measures to express sample relatedness are the mean and spread of the linear predictor \cite{debray2013framework} or the correlation structure of predictors in both samples \cite{kundu2017impact}. The information on sample relatedness can be incorporated in benchmark values of predictive performance to assess model transportability \cite{vergouwe2010external}. While regression coefficients and case-mix distributions clearly quantify sample relatedness, it is impossible to disentangle the sources of discrepancies from these statistical measures. For example, a decrease of the regression coefficients or the spread of the linear predictor at external validation could be due to differences across settings in either person-characteristics or the means by which these characteristics were measured. Moreover, less precise predictor measurements affect both the regression coefficients and the spread of the linear predictor, meaning that measurement heterogeneity can mask similarities and differences between the individuals in a derivation and validation sample. Knowledge of substantive differences between derivation and validation setting can help researchers determining to which extent the prediction model is transportable.\par
In theory, measurement error correction procedures could be applied to adjust for measurement heterogeneity when data on both $X$ and $W$ are available \cite{keogh2014toolkit}. Alternatively, the degree of measurement heterogeneity could be quantified using the residual intraclass correlation (RICC), which expresses the clustering of measurements across physicians or centers \cite{wynants2013screening}. Yet, we expect that the applicability of these methods in correcting for measurement heterogeneity will be limited not only due to the fact that individual patient data of both the derivation and validation set are required, but furthermore because it is infeasible to disentangle measurement parameters from other characteristics of the data. The main contribution of the taxonomy of measurement error models rises from its aptitude to conceptualize measurement heterogeneity across settings in pragmatic terms.\par
The following implications for prediction studies follow from our work. Ideally, prediction models are derived from predictor measurements that resemble measurement procedures in the intended setting of application. Data collection protocols that reduce measurement error to a minimum do not necessarily benefit the performance of the model as the precision of measurements will most likely not be obtained in validation (or application) settings. Deriving a prediction model from these precise measurements could result in miscalibration similar to model overfitting and reduced discrimination and accuracy at external validation. Furthermore, researchers should bear in mind the implications of using a 'readily available dataset' for model derivation or validation as data quality directly affects predictive performance of the model. For instance, validating a model in a clinical trial dataset, in which measurements typically contain minimal measurement error, may increase measures of discrimination and accuracy, yet the model may provide predicted probabilities too close to the event rate due to miscalibration. Another example is the promising use of large routine care datasets for model validation \cite{steyerberg2017poor,riley2016external,cook2015rise}. Predictor measurement procedures may vary greatly within such datasets or differ from the procedures used to collect the data for the derivation study, which could increase the predictor measurement variance to a level that no longer resembles the amount of measurement variance within a clinical setting. Hence, rather than analyzing data because they are available, prediction models should be derived from and validated on datasets collected with measurement procedures that are in widespread use in the intended clinical setting. Finally, it is important to clearly report which measurement procedures were used for derivation or validation of a prediction model. The influential TRIPOD Statement has drawn attention to the importance of reporting measurement procedures \cite{collins2015transparent}. Our findings indicate that descriptions of measurement procedures at model derivation are essential for proper external validation of the model. Likewise, validation studies ideally contain descriptions of deviations from measurements used at derivation, as these may introduce discrepancies in predictive performance.\par
Our study redefines the importance of predictor measurements in the context of prediction research. We highlight heterogeneity in predictor measurement procedures across settings as an important driver of unanticipated predictive performance at external validation. Preventing measurement heterogeneity at the design phase of a prediction study, both in development and validation studies, facilitates interpretation of predictive performance and benefits the transportability of the prediction model.

\newpage
\section*{acknowledgements}
The authors thank B. Rosche for his technical assistance. 

\section*{disclaimer}
All statements in this report, including its findings and conclusions, are solely those of the authors and do not necessarily represent the views of the Patient-Centered Outcomes Research Institute (PCORI), its Board of Governors or Methodology Committee.

\section*{conflict of interest}
The authors declare that they have no conflict of interest.

\section*{data availability statement}
Data sharing is not applicable to this article as no new data were created or analyzed in this study.

\newpage



\newpage
\begin{landscape}
\begin{table}
\caption{Possible sources of measurement heterogeneity in measurements of predictors, illustrated by examples from previously published prediction studies.}
\begin{tabular}{p{.1\linewidth}p{.15\linewidth}p{.65\linewidth}}
\headrow
\textbf{Type of \newline predictor} & \textbf{Examples of predictors} & \textbf{Examples of measurement heterogeneity}\\
Anthropometric measurements & Height \newline Weight \newline Body circumference  & Guidelines on imaging decisions in osteoporosis care are established using standardized measurements of height, while in clinical practice height is measured using non-standardized techniques or self-reported values \cite{mikula2016clinical}.\\
\hline
Physiological\newline measurements & Blood pressure\newline Serum cholesterol \newline HbA1c\newline Fasting glucose & 
In scientific studies, blood pressure is often measured by the average of multiple measurements performed under standardized conditions, while blood pressure measurements in practice deviate from protocol guidelines in various ways due to variability in available time and devices \cite{drawz2017bp}. \\
\hline
Diagnosis & Previous/current \newline disease  & The diagnosis 'hypertension' can be defined as a blood pressure of $ \geq 140/90$ mm Hg (without use of anti-hypertensive therapy) or as the use of anti-hypertensive drugs \cite{genders2012prediction}.\\
\hline
Treatment/\newline Exposure status & Type of drug used \newline Smoking status \newline Dietary intake & The cut-off value for an 'increased length of stay in the hospital' to predict unplanned readmission may depend on the country in which the model is evaluated \cite{aubert2016prospective}.\\
\hline
Imaging & Presence or size of tissue on ultrasound, MRI, CT or FDG PET scans & In scientific studies, review of FDG PET scans may be protocolized or performed by a single experienced nuclear medicine physician, blinded to patient outcome \cite{herder2006clinical}. In routine practice, FDG PET scans may be reviewed under various systematics or by a multi-disciplinary team \cite{al2015risk}.\\
\hline
\end{tabular}
\label{Examples}
\end{table}
\end{landscape}

\newpage
\begin{table}[h]
\scriptsize
\caption{Input parameters for finite sample simulations. Full-factorial simulations for the parameters $\psi$, $\theta$ and $\sigma_{\epsilon}$ resulted in 54 scenarios for the single-predictor model, and 162 scenarios in both the two-predictor model with and the model without a predictor that was measured consistently across settings. An additional 54 scenarios of differential measurement error in the single-predictor model were evaluated, resulting in a total of 432 scenarios.}
\begin{tabular}{lll}
\headrow
 &\multicolumn{2}{l}{\textbf{Factor values}}\\
  Derivation &$\psi_D$ & 0\\
  &$\theta_D$ & 1.0\\
  &$\sigma_{\epsilon(D)}$ & 0.5, 1.0, 2.0\\
  Validation &$\psi_V$ & 0, 0.25\\
  &$\theta_V$ & 0.5, 1.0, 2.0\\
  &$\sigma_{\epsilon(V)}$ & 0.5, 1.0, 2.0\\
  \hline
\end{tabular}
\label{tab: factorial}
\end{table}

\newpage
\begin{table}[h]
\large
\caption{Out-of-sample predictive performance measures under measurement heterogeneity in a single-predictor logistic regression model. Mean c-statistic, median calibration slope, mean calibration-in-the-large and mean Brier score (standard deviation) at external validation of a single-predictor logistic regression model transported from a derivation set (n= $2,000$) where measurement procedures were described by the random measurement error model (Equation \ref{Random ME}) to validation sets (n= $2,000$) with various measurement structures under Equation (\ref{General MEmodel}). Predictive performance measures were averaged over $10,000$ repetitions. All calibration slopes in the derivation set were equal to 1.0 (0.0) and are therefore not reported.}
\resizebox{\textwidth}{!}{
\begin{tabular}{lllllrll}
  \headrow
 &\thead{Measurement structure} & \multicolumn{2}{c}{\textbf{C-statistic}} & \thead{Calibration} & \thead{Calibration-in-} & \multicolumn{2}{c}{\textbf{Brier score}}  \\ 
\headrow
& \thead{at validation} &  \thead{Derivation} & \thead{Validation} & \thead{slope} & \thead{the-large ($\times 10$)} & \thead{Derivation} & \thead{Validation}\\
  $\sigma_{\epsilon(D)}^2 < \sigma_{\epsilon(V)}^2$ & $\psi = 0$, $\theta = 0.5$ & 0.745 (0.033) & 0.590 (0.034) & 0.247 (0.153) & -0.002 (0.006) & 0.204 (0.012) & 0.281 (0.033)\\
  & $\psi = 0$, $\theta = 1.0$ & 0.745 (0.033) & 0.655 (0.045) &  0.380 (0.180) & 0.008 (0.014) &0.204 (0.012) & 0.257 (0.031)\\ 
  & $\psi = 0$, $\theta = 2.0$ & 0.745 (0.033) & 0.726 (0.033) &  0.428 (0.125) & -0.009 (0.003) &  0.204 (0.012) & 0.232 (0.023)\\ 
  & $\psi = 0.25$, $\theta = 0.5$ & 0.745 (0.033) & 0.589 (0.034) &  0.247 (0.153) & -2.202 (0.643) &  0.204 (0.012) & 0.283 (0.032)\\ 
  & $\psi = 0.25$, $\theta = 1.0$ & 0.745 (0.033) & 0.655 (0.045) &  0.380 (0.180) & -2.210 (0.652) & 0.204 (0.012) & 0.258 (0.031)\\ 
  & $\psi = 0.25$, $\theta = 2.0$ & 0.745 (0.033) & 0.726 (0.033) &  0.428 (0.125) & -2.205 (0.651) & 0.204 (0.012) & 0.233 (0.023)\\ 
  $\sigma_{\epsilon(D)}^2 = \sigma_{\epsilon(V)}^2$ & $\psi = 0$, $\theta = 0.5$ & 0.700 (0.068) & 0.635 (0.069) & 0.812 (0.291) & 0.001 (0.006) & 0.217 (0.020) & 0.235 (0.015) \\ 
  & $\psi = 0$, $\theta = 1.0$ & 0.700 (0.068) & 0.700 (0.068) &  1.000 (0.000) & 0.001 (0.008) & 0.217 (0.020) & 0.218 (0.020)\\ 
  & $\psi = 0$, $\theta = 2.0$ & 0.700 (0.068) & 0.753 (0.042) &  0.955 (0.377) & -0.002 (0.013) &  0.217 (0.020) & 0.204 (0.014)\\ 
  & $\psi = 0.25$, $\theta = 0.5$ & 0.700 (0.068) & 0.635 (0.069) &  0.811 (0.293) & -1.529 (1.027) & 0.217 (0.020) & 0.237 (0.014)\\ 
  & $\psi = 0.25$, $\theta = 1.0$ & 0.700 (0.068) & 0.700 (0.068) &  1.002 (0.002) & -1.530 (1.033) & 0.217 (0.020) & 0.219 (0.019)\\ 
  & $\psi = 0.25$, $\theta = 2.0$ & 0.700 (0.068) & 0.753 (0.042) &  0.955 (0.377) & -1.526 (1.024) &  0.217 (0.020) & 0.205 (0.013)\\ 
  $\sigma_{\epsilon(D)}^2 > \sigma_{\epsilon(V)}^2$ & $\psi = 0$, $\theta = 0.5$ & 0.655 (0.045) & 0.681 (0.045) & 3.147 (1.991) &  0.003 (0.007) & 0.230 (0.011) & 0.234 (0.009)\\ 
  & $\psi = 0$, $\theta = 1.0$ & 0.655 (0.045) & 0.745 (0.034)  & 3.106 (1.563) & 0.000 (0.006) &  0.230 (0.011) & 0.220 (0.014)\\ 
  & $\psi = 0$, $\theta = 2.0$ & 0.655 (0.045) & 0.781 (0.014) & 2.160 (0.969) & 0.005 (0.009) &  0.230 (0.011) & 0.203 (0.013)\\ 
  & $\psi = 0.25$, $\theta = 0.5$ & 0.655 (0.045) & 0.681 (0.045) & 3.156 (2.001) & -0.846 (0.528) & 0.230 (0.011) & 0.235 (0.008)\\ 
  & $\psi = 0.25$, $\theta = 1.0$ & 0.655 (0.045) & 0.745 (0.034) & 3.102 (1.559) & -0.846 (0.532) & 0.230 (0.011) & 0.221 (0.013)\\ 
  & $\psi = 0.25$, $\theta = 2.0$ & 0.655 (0.045) & 0.781 (0.014) & 2.159 (0.967) & -0.851 (0.535) & 0.230 (0.011) & 0.203 (0.013)  \\ 
   \hline
\end{tabular}}
\label{M1results}
\end{table}

\newpage
\begin{table}[h]
\caption{Effects of differential measurement of predictors in events and non-events in four scenarios. Mean c-statistic, median calibration slope and mean Brier score (standard deviation) averaged over 10,000 repetitions for a single-predictor logistic regression model under four specific measurement error structures varying in the degree of random measurement variance under the differential measurement error model (Equation \ref{General MEmodel}). By default, $\sigma_{\epsilon}^2$ is set to $1.0$. When $\sigma_{\epsilon1}^2 = 0.5$, measurements are more precise in cases. When $\sigma_{\epsilon1}^2 = 2.0$, measurements are less precise in cases.}
\label{diffresults}
\resizebox{\textwidth}{!}{
\begin{tabular}{lrccccc}
  \headrow
 && \multicolumn{2}{c}{C-statistic}& Calibration & \multicolumn{2}{c}{Brier score}\\
 \headrow
 \multicolumn{2}{l}{Differential measurement error at...}& Derivation & Validation&  slope& Derivation & Validation\\
 Derivation&$\sigma_{\epsilon1}^2 = 0.5$ & 0.730 (0.011) & 0.707 (0.012) & 0.780 (0.071)& 0.209 (0.004) & 0.219 (0.004)\\
 &$\sigma_{\epsilon1}^2 = 2.0$& 0.655 (0.012) & 0.707 (0.012) & 1.856 (0.208) & 0.231 (0.003) & 0.223 (0.002)\\
 Validation& $\sigma_{\epsilon1}^2 = 0.5$ & 0.706 (0.012) & 0.730 (0.011) & 1.293 (0.120) & 0.217 (0.003) & 0.211 (0.003) \\
 & $\sigma_{\epsilon1}^2 = 2.0$ & 0.706 (0.012) & 0.655 (0.012) & 0.547 (0.061) & 0.217 (0.004) & 0.237 (0.005) \\
  \hline
\end{tabular}
}
\end{table}

\newpage 
\begin{table}
\caption{Key Findings. Effects of measurement heterogeneity on predictive performance in general scenarios of measurement heterogeneity. The scenarios were defined by generating different qualities of measurement across settings using the general measurement error model in Equation (\ref{General MEmodel}). Measurements in the derivation set corresponded to the random measurement error model (Equation \ref{General MEmodel}), i.e. under $\psi_D = 0$ and $\theta_D = 1.0$. Using similar logic, all patterns can be translated to differential measurement of cases and non-cases (i.e. when $\psi_1 \neq \psi_0$ and/or $\theta_1 \neq \theta_0$ and/or $\sigma_{\epsilon1}^2 \neq \sigma_{\epsilon0}^2$).}
\resizebox{\textwidth}{!}{
\begin{tabular}{llllll}
\headrow
&&\multicolumn{3}{l}{\textbf{Predictive performance at validation}}&\\
\headrow
\textbf{Predictor measurements at validation}&&\textbf{Discrimination}&\textbf{Calibration-in-the-large}&\textbf{Calibration slope}&\textbf{Overall accuracy}\\
\hiderowcolors
Less precise compared to derivation;&$\sigma_{\epsilon(D)}^2 < \sigma_{\epsilon(V)}^2$&Deteriorated& - & $b <1$&Deteriorated\\
More precise compared to derivation;&$\sigma_{\epsilon(D)}^2 > \sigma_{\epsilon(V)}^2$&Improved& - &$b >1$&Improved\\
Weaker association with actual predictor value, while &&&&\\
\hspace*{3mm} - less precise compared to derivation;&$\theta_V < 1.0$,\hspace*{1mm} $\sigma_{\epsilon(D)}^2 < \sigma_{\epsilon(V)}^2$&Stronger deterioration&-&Stronger $b <1$&Stronger deterioration\\
\hspace*{3mm} - more precise compared to derivation;&$\theta_V < 1.0$,\hspace*{1mm} $\sigma_{\epsilon(D)}^2 > \sigma_{\epsilon(V)}^2$&Less improvement&-&Stronger $b>1$&Less improvement\\
Stronger association with actual predictor value, while &&&&\\
\hspace*{3mm} - less precise compared to derivation;&$\theta_V > 1.0$,\hspace*{1mm} $\sigma_{\epsilon(D)}^2 < \sigma_{\epsilon(V)}^2$&Less deterioration&-&Less $b <1$&Less deterioration\\
\hspace*{3mm} - more precise compared to derivation;&$\theta_V > 1.0$,\hspace*{1mm} $\sigma_{\epsilon(D)}^2 > \sigma_{\epsilon(V)}^2$&Stronger improvement&-&Less $b>1$&Stronger improvement\\
Increased by a constant relative to derivation.&$\psi_V > 0$& - & $a < 0$ & - & - \\
\hline
\end{tabular}}
\end{table}

\clearpage
\pagenumbering{gobble}
\begin{figure}[t]
\resizebox{\textwidth}{!}{\includegraphics[]{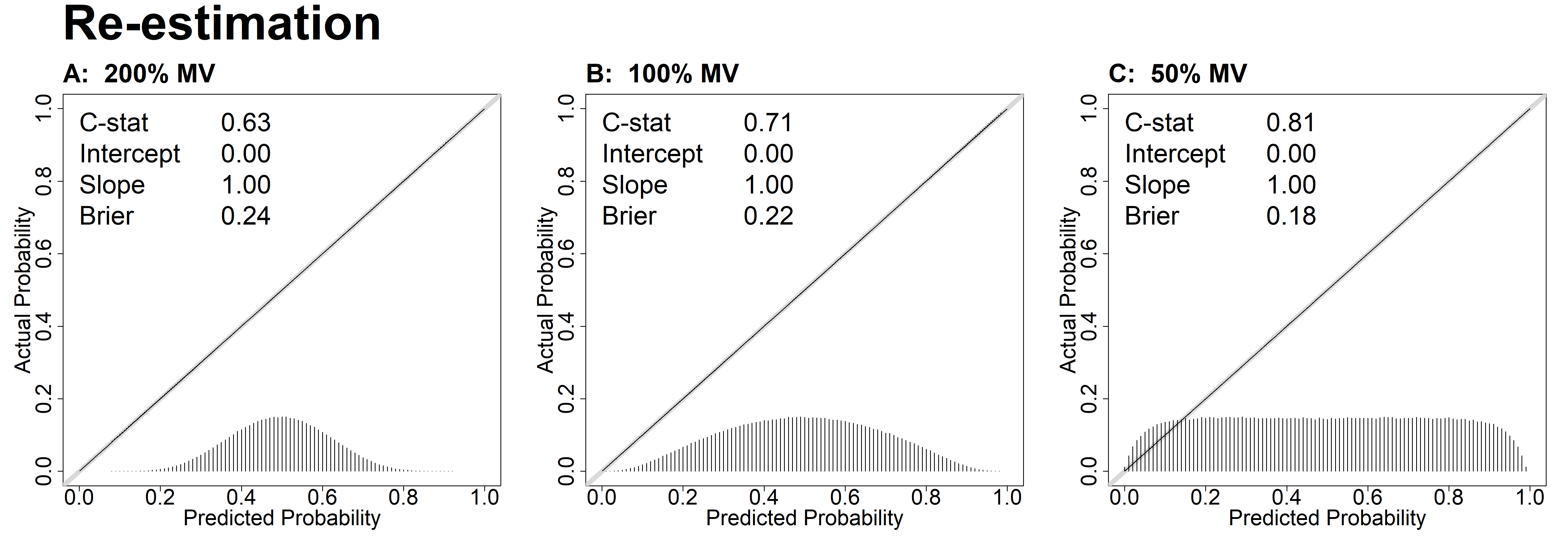}}\\
\vspace{2mm}
\resizebox{\textwidth}{!}{\includegraphics[]{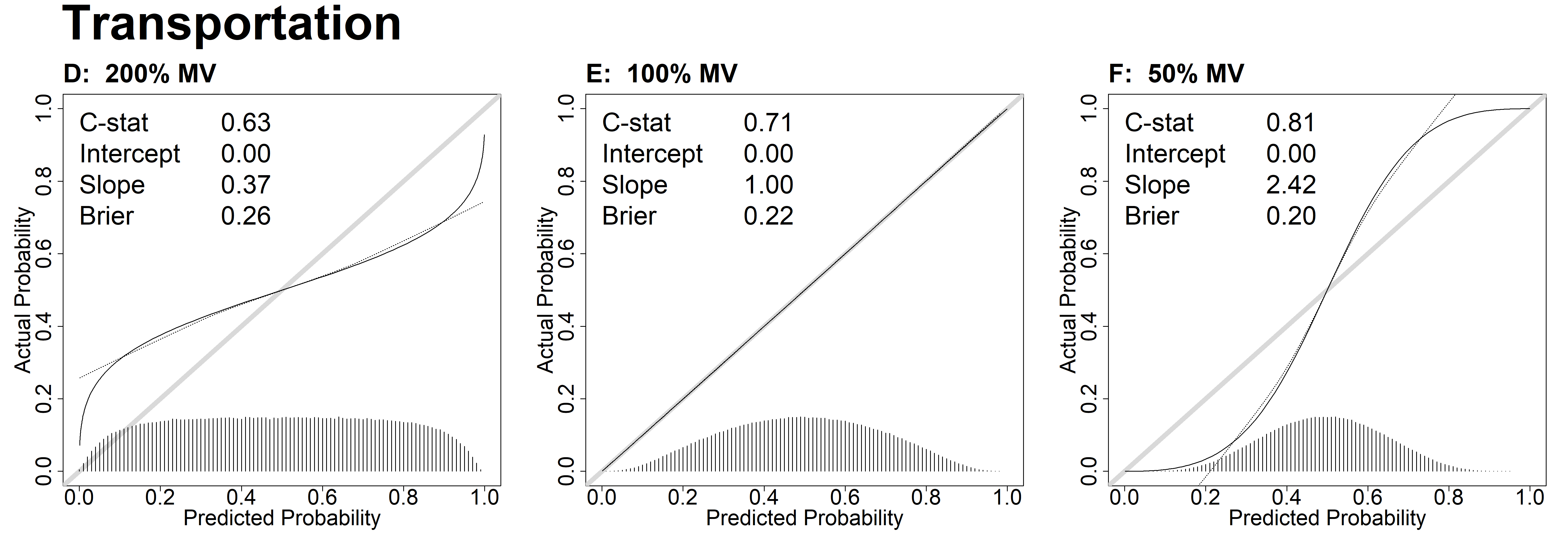}}\\
\caption{\textbf{Measures of predictive performance under predictor measurement error and predictor measurement heterogeneity.}\newline MV = measurement variance of the predictor measurement used for model validation relative to the predictor measurement used for derivation. The data generating mechanism corresponded perfectly to the estimated logistic regression model. The top rows show calibration plots of a single-predictor model that is fitted using predictor measurement $x$ and validated by re-estimating the model on the same data using $w$. The bottom rows show situations where the same model is transported from derivation to validation setting, specifically, the model (D) is derived using $x$ and validated using $w$, (E) is derived and validated using $w$, and (F) is derived using $w$ and validated using $x$. The calibration plots show the calibration slope (black line) and predicted probability frequencies (bottom-histograms) for situations in which the predictor measurement variance at validation equals 200\% (A,D), 100\% (B,E), or 50\% (C,F) of the predictor measurement variance at derivation. The \texttt{val.prob} function from the rms package was used to compute the simulation outcome measures and to generate the calibration plots \cite{harrell2013rms}, where we edited the legend format settings in the plot to improve readability.}
\end{figure}

\newpage
\begin{figure}[h]
	\centering
\resizebox{\textwidth}{!}{\includegraphics[]{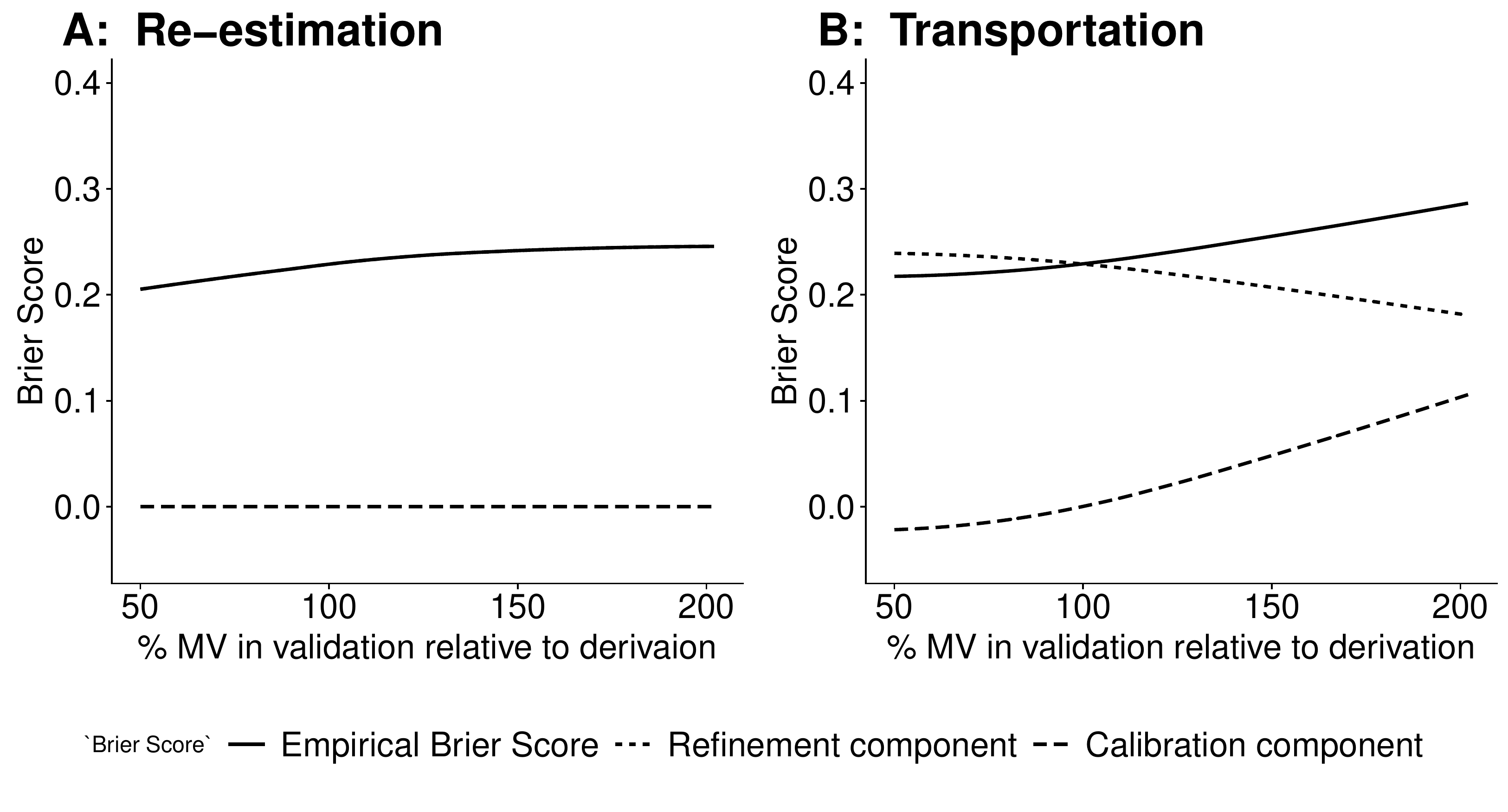}}
\caption{\textbf{Decomposed Brier score under predictor measurement error and predictor measurement heterogeneity.}\newline MV = measurement variance of the predictor measurement used for model validation relative to the predictor measurement used for derivation. The data generating mechanism corresponded perfectly to the estimated logistic regression model. The plot displays the large sample properties of the components of the Brier score (Equation \ref{Decomposed Brier}) under increasing random predictor measurement variance at validation, corresponding to the random measurement error model (Equation \ref{Random ME}). The left panel shows the Brier score for a single-predictor logistic regression model that is fitted using predictor measurement $x$ and validated by re-estimating the model on the same data using $w$. The right panel shows transportation from $w$ at derivation to $x$ at validation (upto \%MV = 100) and transportation from $x$ at derivation to $w$ at validation (from \%MV = 100 onwards).}
\end{figure}

\clearpage
\begin{figure}[h]
\label{M2results}
\centering
\includegraphics[width=\textwidth]{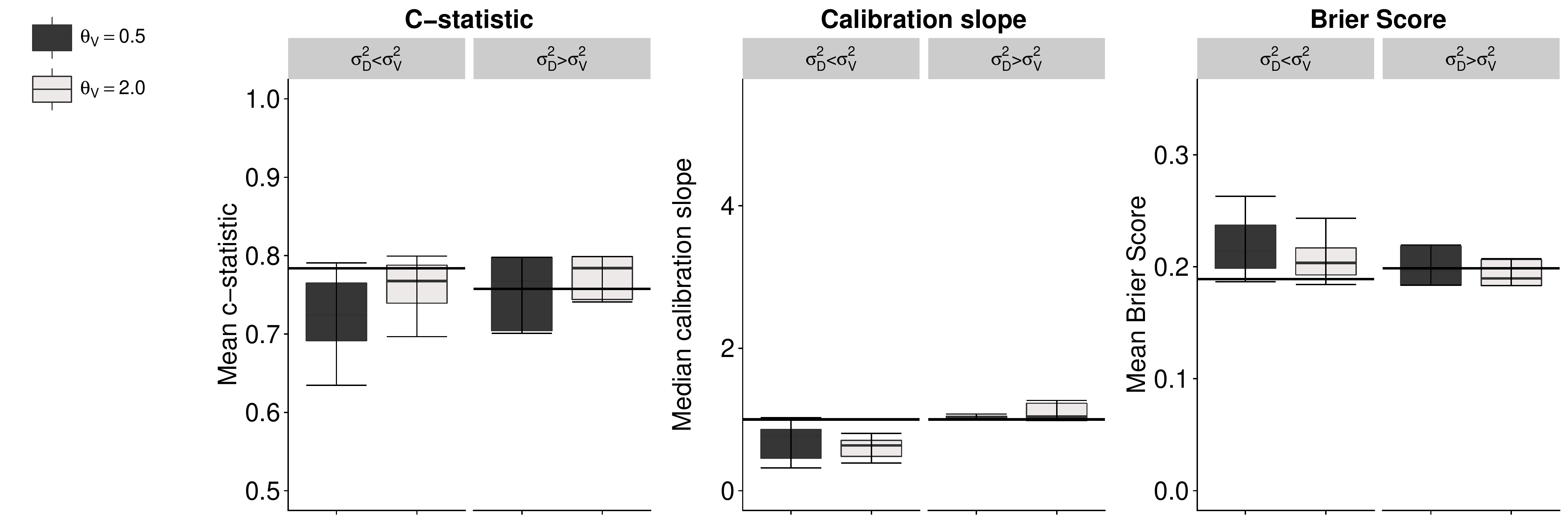}
\caption{\textbf{Measures of predictive performance under measurement heterogeneity in one of two predictors in finite sample simulations.}\newline Mean c-statistic, median calibration slope and mean Brier score averaged over $10,000$ repetitions with interquartile range and 95\% confidence interval for a two-predictor model where one of the predictors is measured consistent across settings, while the other is measured heterogeneously. Horizontal bars indicate performance measures at model derivation, while boxes indicate performance at external validation. The predictor measurement structure in the derivation set (n= $2,000$) corresponds to the random measurement error model (Equation \ref{Random ME}). In the validation set (n= $2,000$), predictor measurements consist of varying structures under Equation (\ref{General MEmodel}).}
\end{figure}

\newpage
\begin{figure}[h]
\label{M3results}
\centering
\includegraphics[width=\textwidth]{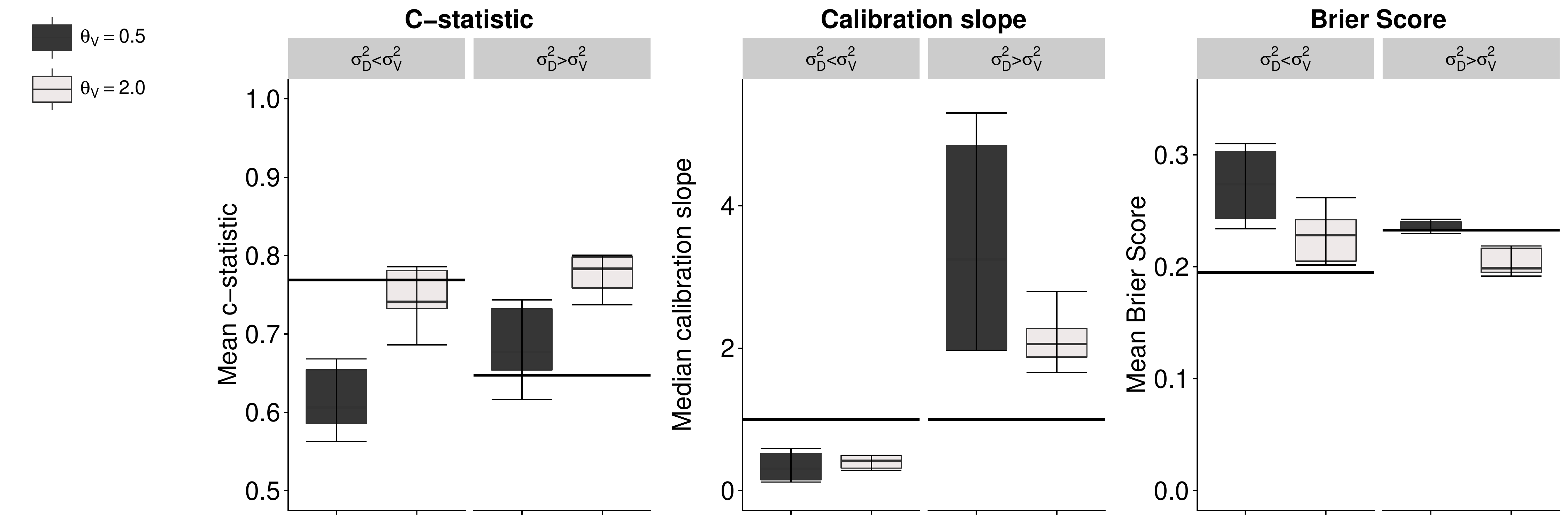}
\caption{\textbf{Measures of predictive performance under measurement heterogeneity in both predictors in finite sample simulations.}\newline Mean c-statistic, median calibration slope and mean Brier score averaged over $10,000$ repetitions with interquartile range and 95\% confidence interval of a two-predictor logistic regression model in which both predictors are measured heterogeneously across settings. Horizontal bars indicate performance measures at model derivation, while boxes indicate performance at external validation. Measurements in the derivation set (n= $2,000$) are recreated using Equation (\ref{Random ME}), which corresponds to the random measurement error model. In the validation set (n= $2,000$), measurements correspond to various measurement structures under Equation (\ref{General MEmodel}).}
\end{figure}

\clearpage
\section*{Appendix 1}

In this appendix, the general effects of measurement heterogeneity on external predictive performance are illustrated in large sample simulations ($N = 1,000,000$). 

\subsection*{Simulation design}
We examined the predictive performance of a single-predictor binary logistic regression model. The data were generated from 
\begin{align*}
\text{logit}(Y)  &= \text{log}(8) X,\\
\text{where } X &\sim \mathcal{N}(0,0.5), 
\end{align*}
and where $X$ reflects the true (often unobserved) underlying value of the predictor. The dataset contained two measurements of the predictor $x$, which were recreated under the general measurement error model (Equation \ref{General MEmodel}). The first measurement, denoted $w_D$, was used to derive the logistic regression model and corresponded to the random error model (Equation \ref{Random ME}). The other measurement, $w_V$, was used to validate the model and corresponded to various measurement structures under the general measurement error model. This validation procedure implies that the model is validated in its original sample, hence, in absence of all other impacts on model transportability. The \texttt{val.prob} function from the rms package in R was used to compute the simulation outcome measures and to generate the calibration plots \cite{harrell2013rms}, where we edited the legend format settings in the plot to improve readability.

\subsection*{Simulation results}
In line with expectations, the predictive performance at validation corresponded perfectly to the predictive performance at derivation when the predictor was measured consistently over settings. The impact on predictive performance when measurements were heterogeneous is described below. 

\subsubsection*{Random measurement heterogeneity}
\begin{figure}[h]
\caption*{\paragraph*{\large{Random measurement heterogeneity}}}
\centering
\resizebox{\textwidth}{!}{\includegraphics[]{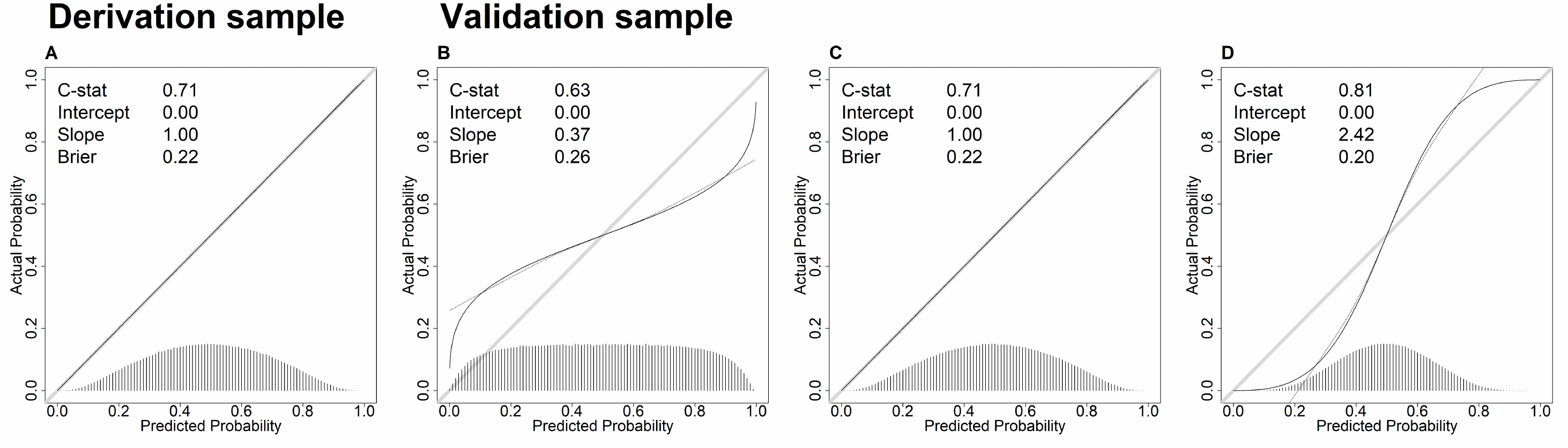}}
\vspace{2mm}
\caption{Predictive performance of a single-predictor binary logistic regression model. The predictor measurement structure corresponds to:}
\resizebox{\textwidth}{!}{\begin{tabular}{lll}
A. $W_D = X + \epsilon_D$,& where $X \sim \mathcal{N}(0,0.5)$ and $\epsilon_D \sim \mathcal{N}(0,0.5)$.&\\
B. $W_V = X + \epsilon_V$,& where $X \sim \mathcal{N}(0,0.5)$ and $\epsilon_V \sim \mathcal{N}(0,2.0)$. &Measurements are less precise at validation, i.e. $\sigma_{\epsilon(D)}^2 < \sigma_{\epsilon(V)}^2$.\\
C. $W_V = X + \epsilon_V$,& where $X \sim \mathcal{N}(0,0.5)$ and $\epsilon_V \sim \mathcal{N}(0,0.5)$. &Measurements consistent across settings, i.e. $\sigma_{\epsilon(D)}^2 = \sigma_{\epsilon(V)}^2$.\\
D. $W_V = X$,& where $X \sim \mathcal{N}(0,0.5)$. &Measurements are more precise at validation, i.e. $\sigma_{\epsilon(D)}^2 > \sigma_{\epsilon(V)}^2$.
\end{tabular}}
\label{random mh}
\end{figure}

When the measurement at validation, in $w_V$, was less precise than at derivation, in $w_D$, i.e. when $\sigma_{\epsilon(D)}^2 < \sigma_{\epsilon(V)}^2$, the c-statistic decreased from $0.71$ at derivation to $0.63$ at validation and the Brier score increased from $0.22$ to $0.26$, indicating a loss in discriminatory power and accuracy. Furthermore, the calibration slope was $0.37$, similar to statistical overfitting (Figure \ref{random mh}b). When the measurement $w_V$ was more precise than $w_D$, i.e. when $\sigma_{\epsilon(D)}^2 > \sigma_{\epsilon(V)}^2$, the c-statistic increased from $0.71$ to $0.81$, and the Brier score decreased from $0.22$ to $0.20$. However, the improved c-statistic and Brier score were accompanied by a calibration slope of $b=2.42$, similar to statistical underfitting (Figure \ref{random mh}d). Calibration-in-the-large was not affected by random measurement heterogeneity.

\subsubsection*{Systematic measurement heterogeneity}
\begin{figure}[h]
\caption*{\paragraph*{\large{Additive systematic measurement heterogeneity}}}
\resizebox{0.67\textwidth}{!}{\includegraphics[]{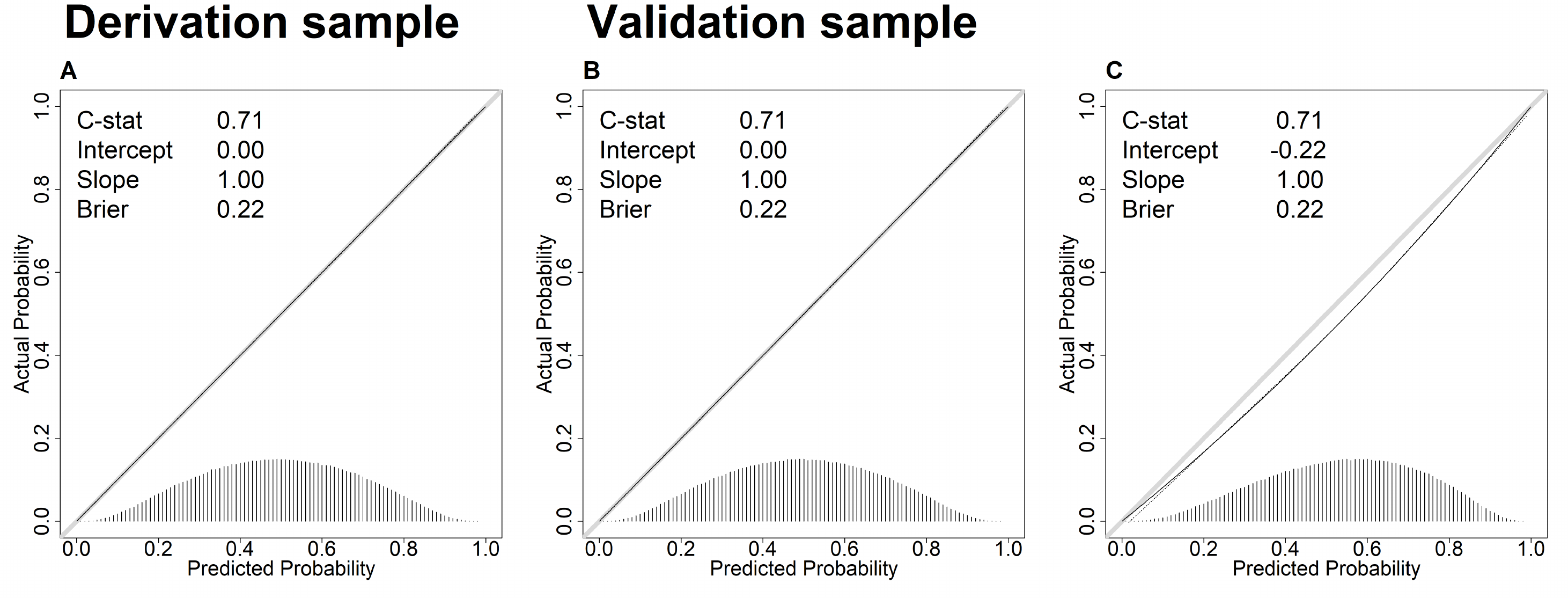}}
\vspace{2mm}
\caption{Predictive performance of a single-predictor binary logistic regression model. The degree of additive error in the validation predictor measurement $W_V$ varies, while the degree of random error is consistent across settings, i.e. while $\sigma_{\epsilon(D)}^2 = \sigma_{\epsilon(V)}^2$. The predictor measurement structure corresponds to:}
\resizebox{\textwidth}{!}{\begin{tabular}{lll}
A. $W_D = X + \epsilon_D$,& where $X \sim \mathcal{N}(0,0.5)$ and $\epsilon_D \sim \mathcal{N}(0,0.5)$.\\
B. $W_V = \psi_V + X + \epsilon_V$,& where $\psi_V = 0$, $X \sim \mathcal{N}(0,0.5)$ and $\epsilon_V \sim \mathcal{N}(0,0.5)$.& Measurements are equal across settings.\\
C. $W_V = \psi_V + X + \epsilon_V$,& where $\psi_V = 0.25$, $X \sim \mathcal{N}(0,0.5)$ and $\epsilon_V \sim \mathcal{N}(0,0.5)$.& Measurements are shifted from $X$ by a constant.
\end{tabular}}
\label{addsysmh}
\end{figure}

\begin{figure}[h]
\caption*{\paragraph*{\large{Multiplicative systematic measurement heterogeneity}}}
\centering
\resizebox{\textwidth}{!}{\includegraphics[]{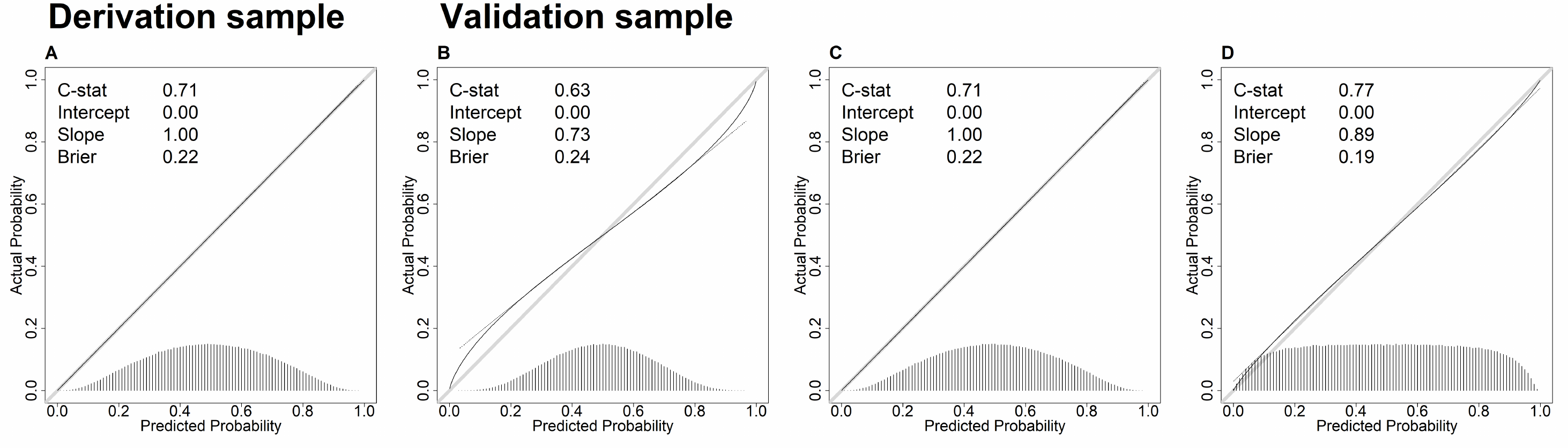}}
\vspace{2mm}
\caption{Predictive performance of a single-predictor binary logistic regression model. The degree of multiplicative error in the validation predictor measurement $W_V$ varies, while the degree of random error is consistent across settings, i.e. while $\sigma_{\epsilon(D)}^2 = \sigma_{\epsilon(V)}^2$. The predictor measurement structure corresponds to:}
\resizebox{\textwidth}{!}{\begin{tabular}{lll}
A. $W_D = \theta_D X + \epsilon_D$,& where $\theta_D = 1.0$, $X \sim \mathcal{N}(0,0.5)$ and $\epsilon_D \sim \mathcal{N}(0,0.5)$.&\\
B. $W_V = \theta_V X + \epsilon_V$,& where $\theta_V = 0.5$, $X \sim \mathcal{N}(0,0.5)$ and $\epsilon_V \sim \mathcal{N}(0,0.5)$.& The association $X$-$W$ is weaker at validation.\\
C. $W_V = \theta_V X + \epsilon_V$,& where $\theta_V = 1.0$, $X \sim \mathcal{N}(0,0.5)$ and $\epsilon_V \sim \mathcal{N}(0,0.5)$.& The association $X$-$W$ is equal across settings.\\
D. $W_V = \theta_V X + \epsilon_V$,& where $\theta_V = 2.0$, $X \sim \mathcal{N}(0,0.5)$ and $\epsilon_V \sim \mathcal{N}(0,0.5)$.& The association $X$-$W$ is stronger at validation.
\end{tabular}}
\label{multsysmh}
\end{figure}

Additive systematic measurement heterogeneity, i.e. $\psi_D \neq \psi_V$, resulted in systematic overestimation of the outcome, which is reflected in the negative value for calibration-in-the-large coefficient, $-0.22$ (Figure \ref{addsysmh}c). Changes in $\psi$ had no apparent effect on the calibration slope, c-statistic, and Brier score. Multiplicative systematic measurement heterogeneity at validation, in $w_V$, i.e. $\theta_V \neq 1$, in combination with random measurement error led to a calibration slope $b<1$. The impact on the c-statistic and the Brier score was in the direction of association between $x$ and $w$. When this association was relatively weak, e.g. when $\theta_V = 0.5$, the c-statistic decreased from $0.71$ to $0.63$ and the Brier score increased from $0.22$ to $0.24$ (Figure \ref{multsysmh}b). When the association between $x$ and $w_V$ was relatively strong, e.g. when $\theta_V = 2.0$, the c-statistic improved from $0.71$ to $0.77$ and the Brier score improved from $0.22$ to $0.19$ (Figure \ref{multsysmh}d).

\subsubsection*{Differential measurement heterogeneity}
\begin{figure}[h]
\caption*{\paragraph*{\large{Differential measurement heterogeneity at validation}}}
\resizebox{0.5\textwidth}{!}{\includegraphics[]{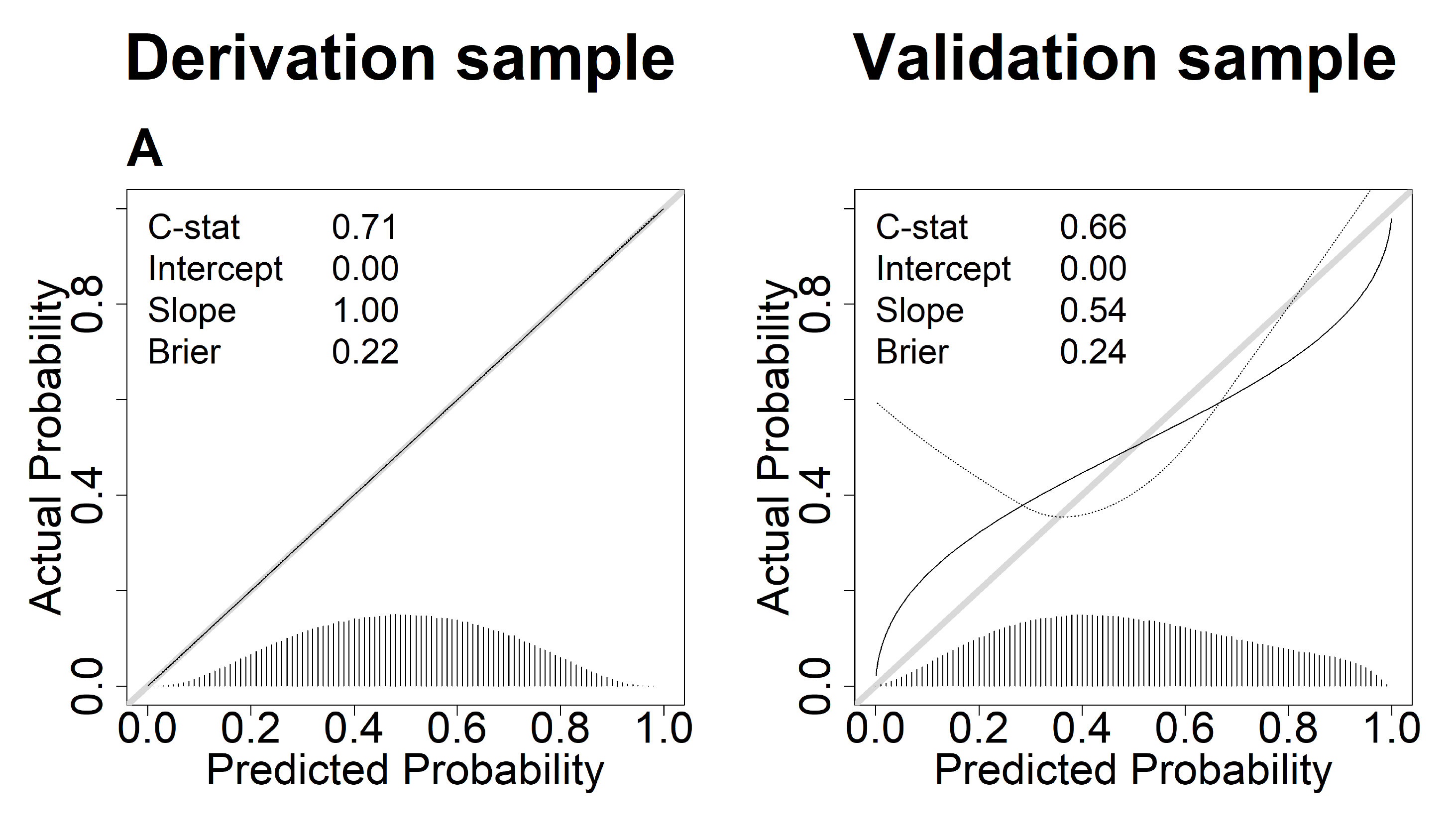}}\\
\resizebox{0.5\textwidth}{!}{\includegraphics[]{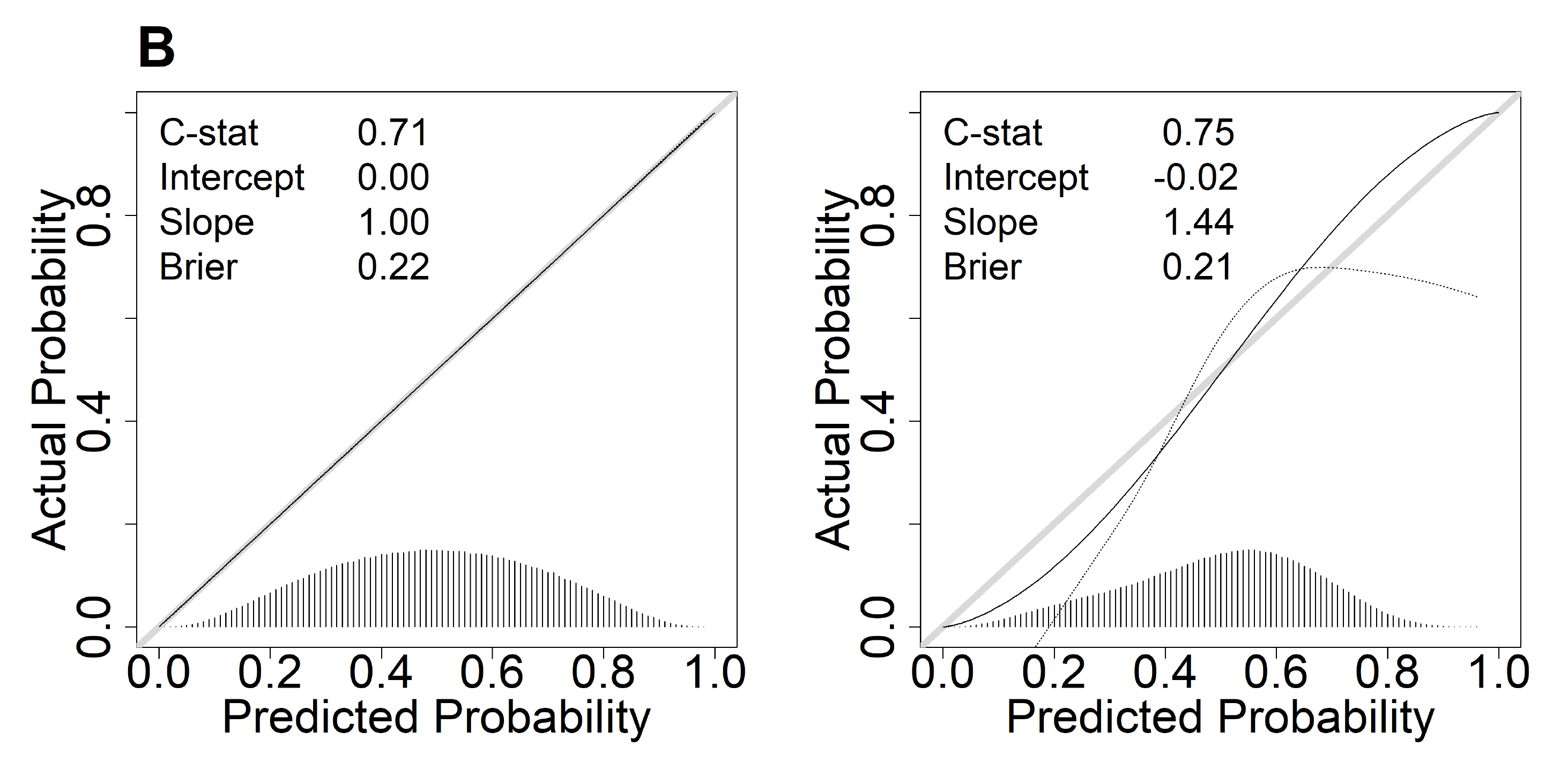}}\\
\resizebox{0.5\textwidth}{!}{\includegraphics[]{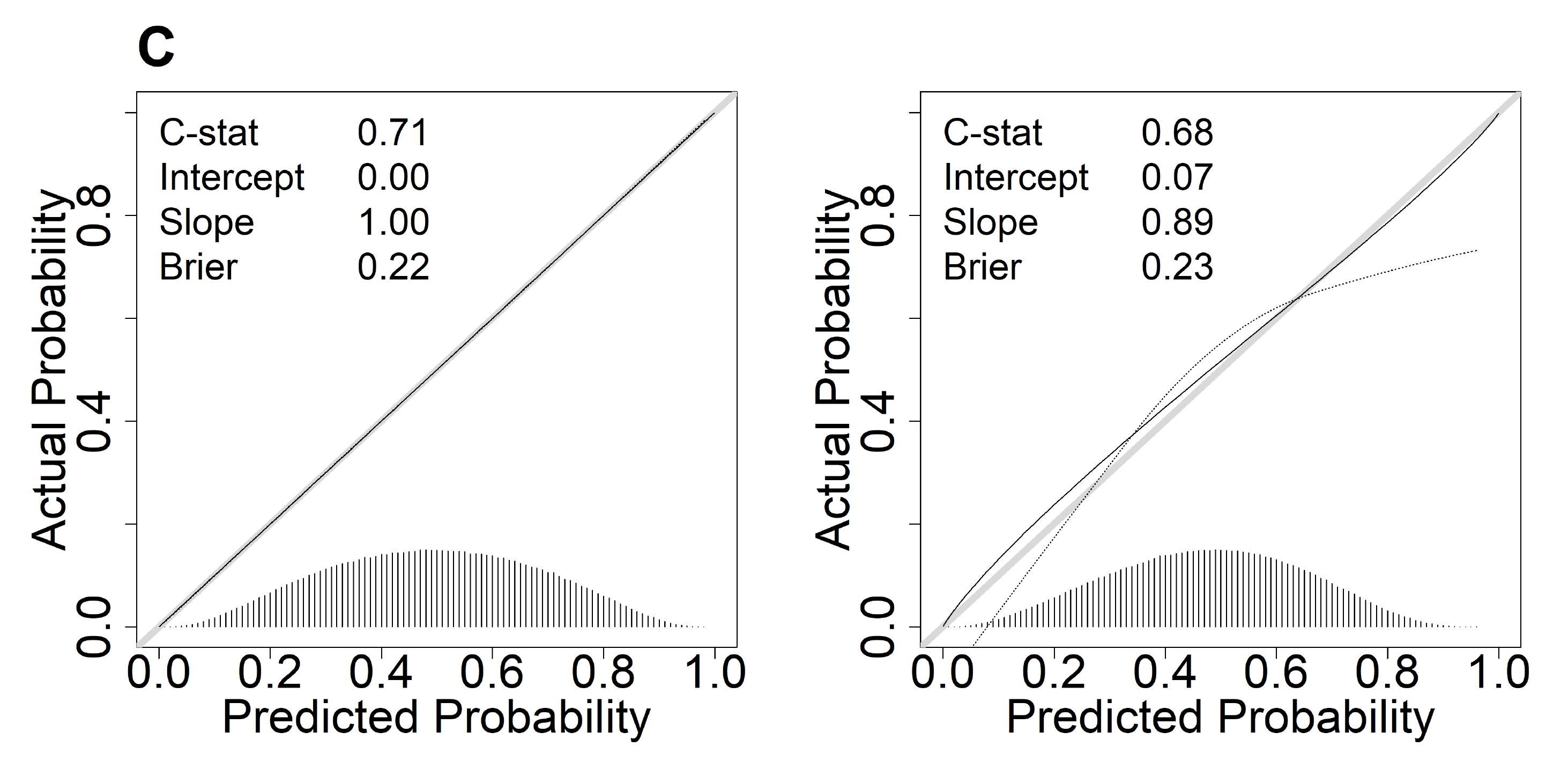}}
\vspace{2mm}
\caption{Predictive performance of a single-predictor binary logistic regression model. In all three scenarios, $\psi_{D\{0,1\}}$ and $\psi_{V\{0,1\}}$ equal 0, the default value for $\theta_{D\{0,1\}}$ and $\theta_{V\{0,1\}}$ is $1.0$, and the default value for $\sigma_{\epsilon(D\{0,1\})}^2$ and $\sigma_{\epsilon(V\{0,1\})}^2$ is $0.5$. Otherwise, the predictor measurement structure for the cases at validation (specified by $\theta_{V1}$ and $\sigma_{\epsilon(V1)}^2$) corresponds to:}
\resizebox{\textwidth}{!}{\begin{tabular}{lll}
A. $W_V = \theta_V X + \epsilon_V$,& where $X \sim \mathcal{N}(0,0.5)$ and $\epsilon_{V1} \sim \mathcal{N}(0,2.0)$.& Measurements of cases are less precise at validation.\\
B. $W_V = \theta_V X + \epsilon_V$,& where $X \sim \mathcal{N}(0,0.5)$ and $\epsilon_{V1} \sim \mathcal{N}(0,0)$.& Measurements of cases are more precise at validation.\\
C. $W_V = \theta_V X + \epsilon_V$,& where $\theta_{V1} = 0.5$, $X \sim \mathcal{N}(0,0.5)$ and $\epsilon_{V} \sim \mathcal{N}(0,0.5)$.& Associations between $X$ and $W$ in cases are weaker at validation.\\
\end{tabular}}
\label{diffmhv}
\end{figure}

All forms of differential measurement of cases and non-cases led to miscalibration at external validation. For example, when measurement of cases was less precise at validation, in $w_V$, i.e. $\sigma_{\epsilon1(V)}^2 > \sigma_{\epsilon0(V)}^2$, the calibration slope at validation was $0.54$. The c-statistic decreased from $0.71$ to $0.66$, the Brier score increased from $0.22$ to $0.24$ (Figure \ref{diffmhv}a). In case of systematic differential measurement of cases and non-cases, when the association between $x$ and $w$ in cases was weaker in $w_V$, i.e. $\theta_{1D} > \theta_{1V}$, the c-statistic decreased from $0.71$ to $0.68$, the Brier score increased from $0.22$ to $0.23$, and the calibration slope was $0.89$ (Figure \ref{diffmhv}c).\par
\begin{figure}[h]
\caption*{\paragraph*{\large{Differential measurement heterogeneity at derivation}}}
\resizebox{0.5\textwidth}{!}{\includegraphics[]{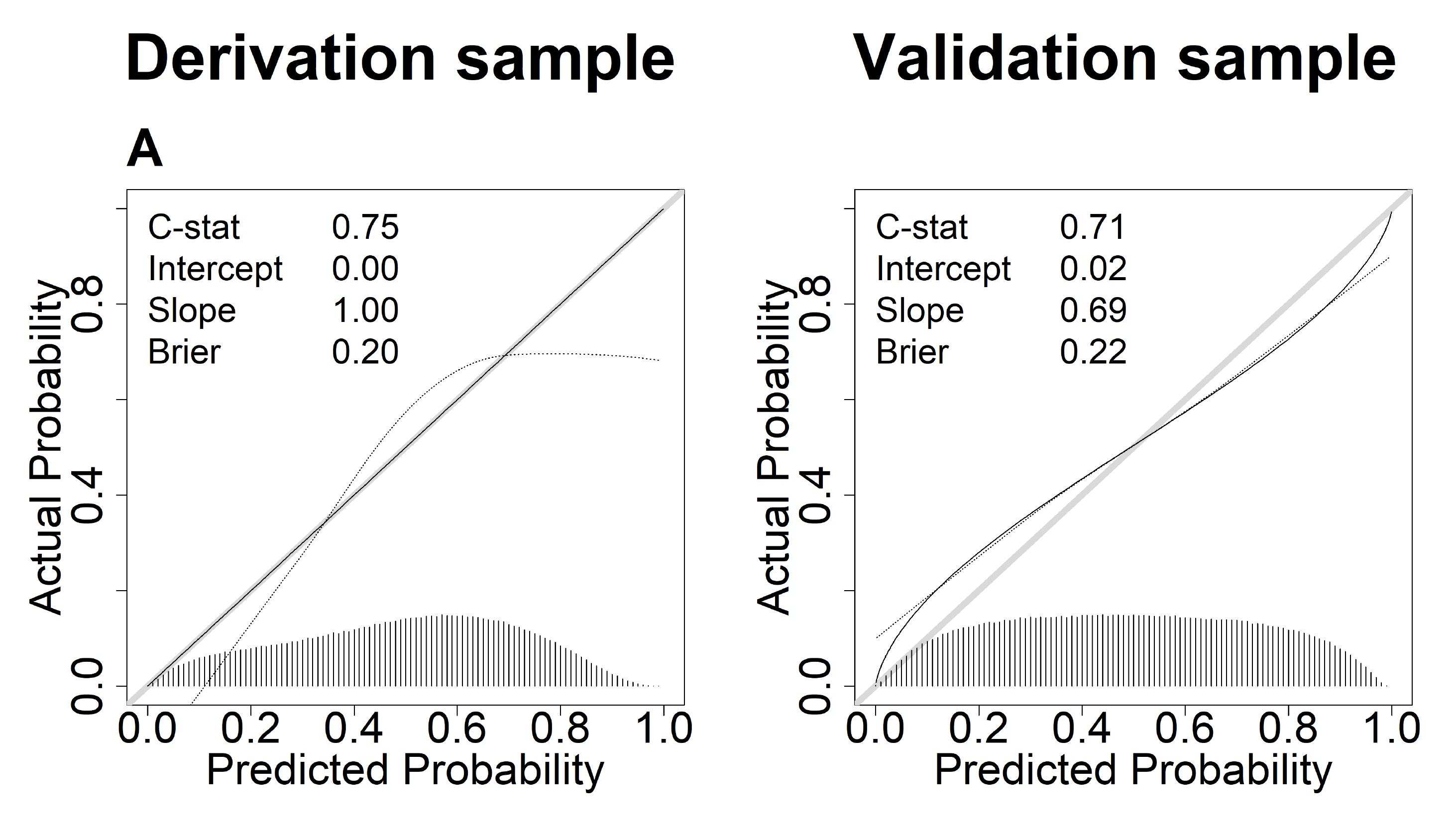}}\\
\resizebox{0.5\textwidth}{!}{\includegraphics[]{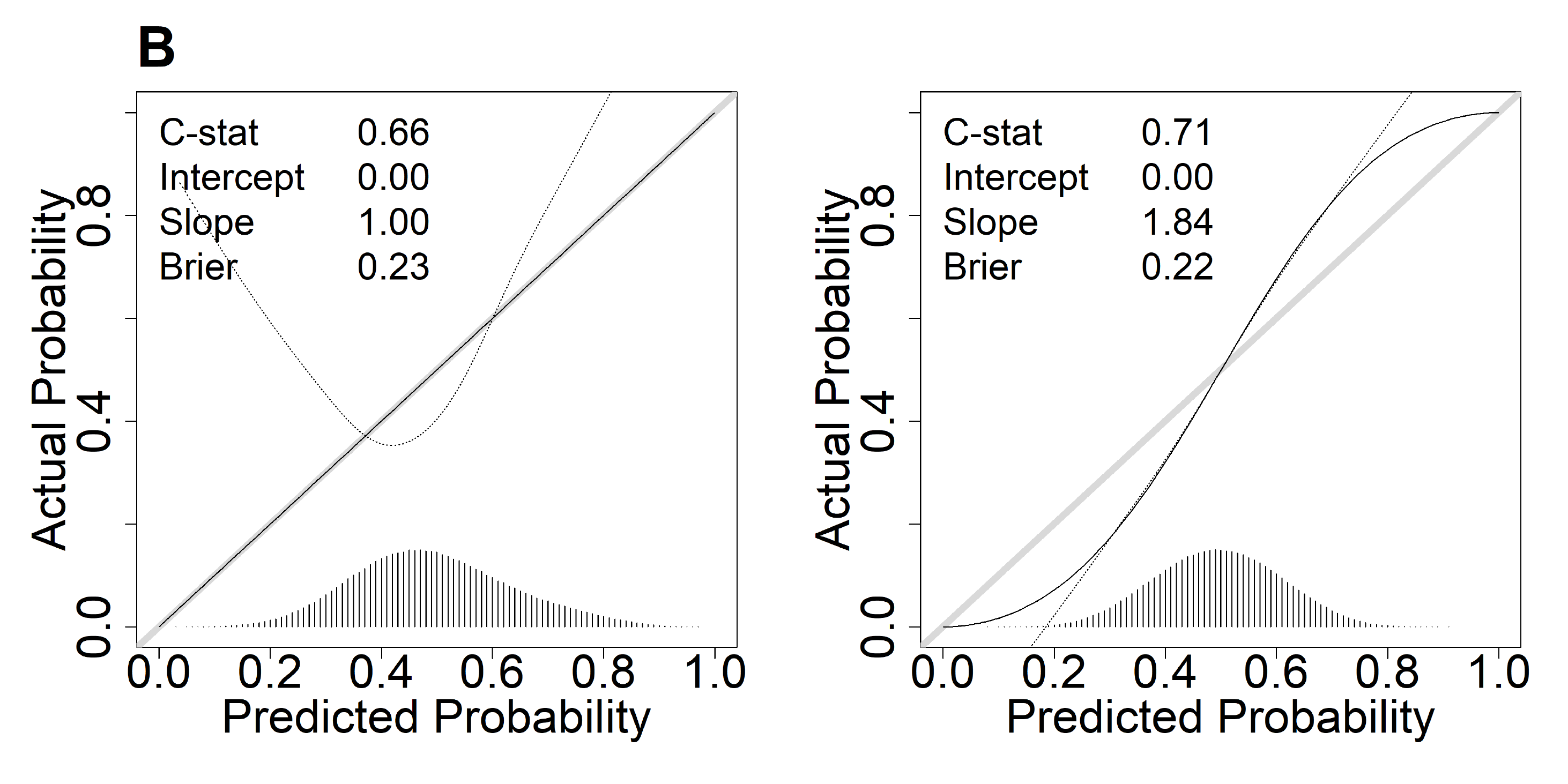}}\\
\resizebox{0.5\textwidth}{!}{\includegraphics[]{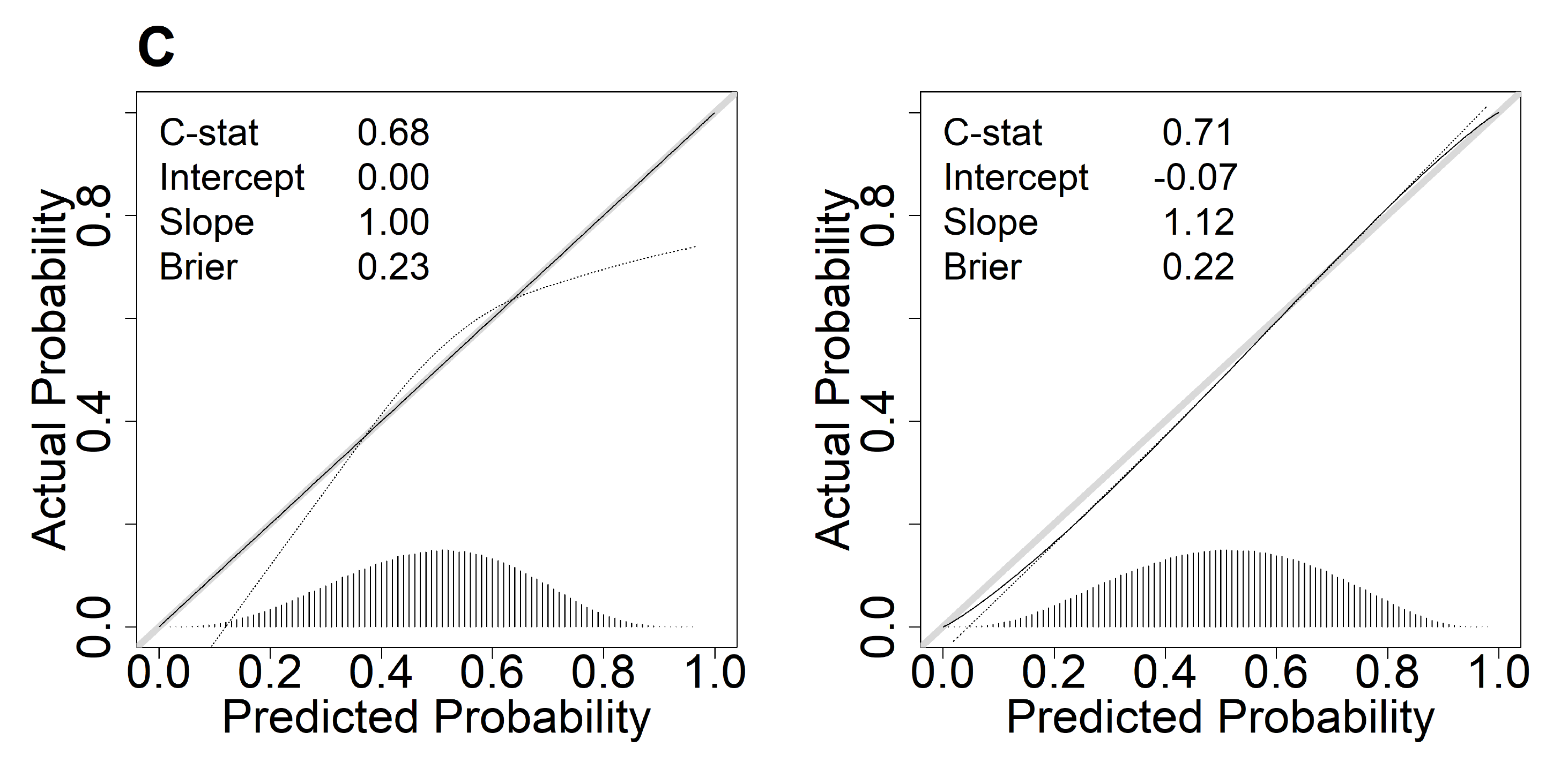}}
\vspace{2mm}
\caption{Predictive performance of a single-predictor binary logistic regression model. In all three scenarios, $\psi_{D\{0,1\}}$ and $\psi_{V\{0,1\}}$ equal 0, the default value for $\theta_{D\{0,1\}}$ and $\theta_{V\{0,1\}}$ is $1.0$, and the default value for $\sigma_{\epsilon(D\{0,1\})}^2$ and $\sigma_{\epsilon(V\{0,1\})}^2$ is $0.5$. Otherwise, the predictor measurement structure for the cases at derivation (specified by $\theta_{D1}$ and $\sigma_{\epsilon(D1)}^2$) corresponds to:}
\resizebox{\textwidth}{!}{\begin{tabular}{lll}
A. $W_D = \theta_D X + \epsilon_D$,& where $X \sim \mathcal{N}(0,0.5)$ and $\epsilon_{D1} \sim \mathcal{N}(0,0)$.& Measurements of cases are more precise at derivation.\\
B. $W_D = \theta_D X + \epsilon_D$,& where $X \sim \mathcal{N}(0,0.5)$ and $\epsilon_{D1} \sim \mathcal{N}(0,2.0)$.& Measurements of cases are less precise at derivation.\\
C. $W_D = \theta_D X + \epsilon_D$,& where $\theta_{D1} = 0.5$, $X \sim \mathcal{N}(0,0.5)$ and $\epsilon_{D} \sim \mathcal{N}(0,0.5)$.& Associations between $X$ and $W$ in cases are weaker at derivation.
\end{tabular}}
\label{diffmhd}
\end{figure}

Inverse effects on predictive performance were found when cases and non-cases were measured differentially at derivation, in $w_D$. When measurement of cases was less precise at derivation, i.e. $\sigma_{\epsilon1(D)}^2 > \sigma_{\epsilon0(D)}^2$, the c-statistic increased from $0.66$ to $0.71$, the Brier score decreased from $0.23$ to $0.22$, and the calibration slope at validation was $1.84$ (Figure \ref{diffmhd}b). When the association between $x$ and $w$ in cases was weaker at derivation, in $w_D$, i.e. $\theta_{1D} < \theta_{1V}$, the c-statistic improved from $0.68$ to $0.71$, the Brier score improved from $0.23$ to $0.22$, and the calibration slope was $1.12$ (Figure \ref{diffmhd}c).

\newpage

\newpage

\newpage

\newpage

\newpage

\clearpage 
\noindent\section*{Supplementary Figure 1}
\begin{figure}[h!]
\includegraphics[]{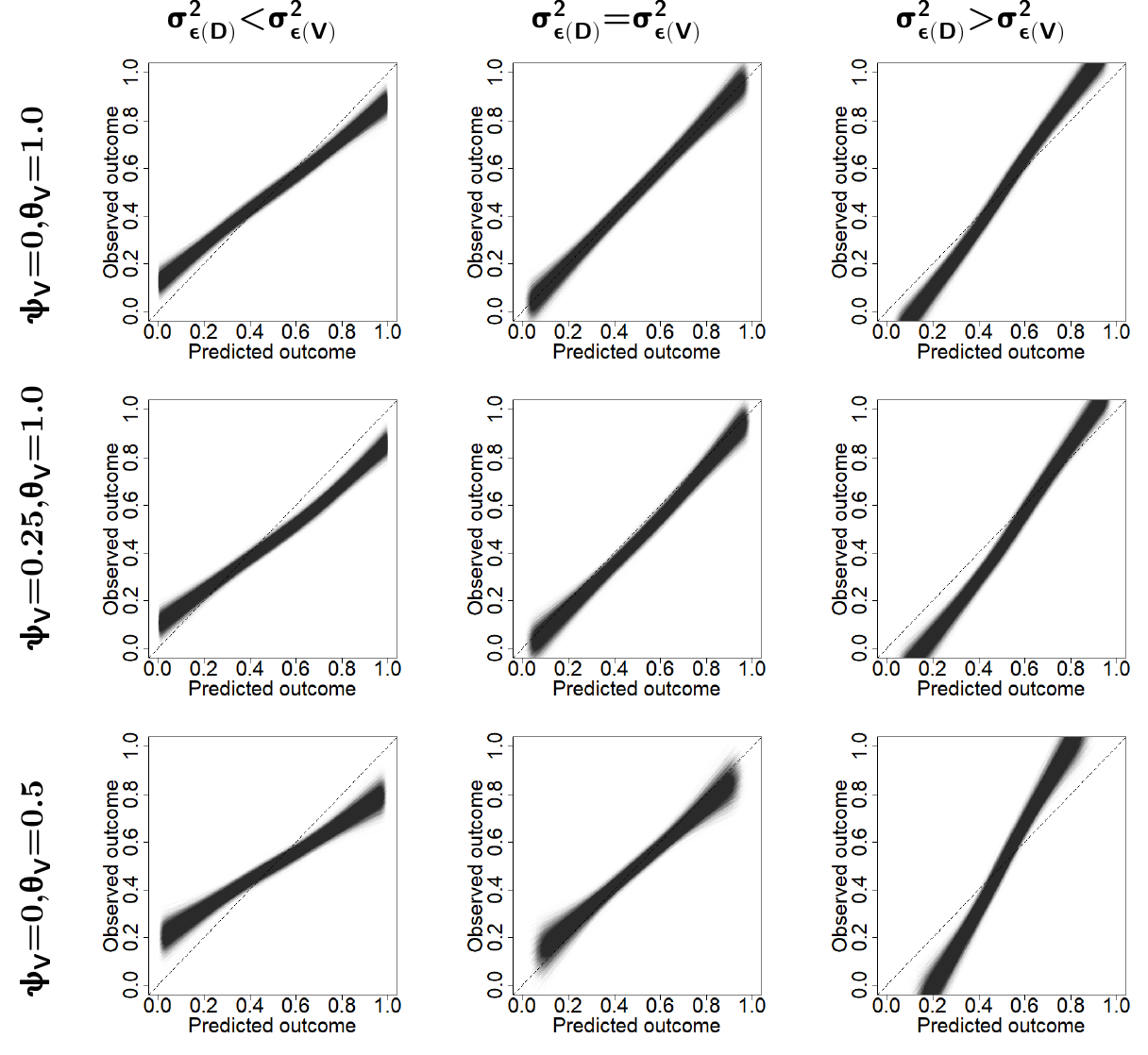}
\caption*{Lowess calibration curves are overlaid for 10,000 resamplings for 9 scenarios of predictor measurement heterogeneity in the two-predictor model in which both predictors are measured heterogeneously. The figure titles indicate the parameters of the general measurement error model (Equation \ref{General MEmodel}) to which the predictor measurements at validation correspond.}
\end{figure}

\end{document}